\documentclass[12pt]{iopart}
\usepackage{graphicx,iopams}
\usepackage{esint}
\usepackage{graphicx}
\usepackage{epsfig}
\usepackage{subfig}
\usepackage{pstricks}
\usepackage{amssymb}
\usepackage{color}
\usepackage{psfrag}
\definecolor{dblue}{rgb}{0.1,0.1,0.44}
\definecolor{dgreen}{rgb}{0.2 ,0.54, 0.2}
\newcommand{\be}{\begin{equation}}
\newcommand{\ee}{\end{equation}}
\newcommand{\bea}{\begin{eqnarray}}
\newcommand{\eea}{\end{eqnarray}}

\begin{document}
     \bibliographystyle{unsrt} 
\title{On the ground-state energy of the finite sine-Gordon ring}
\author{Sergei B.  Rutkevich }
\address{Fakult\"at f\"ur Mathematik und Naturwissenschaften, Bergische Universit\"at Wuppertal, 42097 Wuppertal, Germany.}
\ead{rutkevich@uni-wuppertal.de}
\begin{abstract}
The Casimir scaling function  characterising  the ground-state energy of the sine-Gordon model
in a finite circle has been studied  analytically and numerically both in the repulsive and attractive
regimes.  The numerical calculations of the scaling function at several values of the coupling constant were performed by the iterative solution of the Destri-de Vega nonlinear integral equations. 
The ultraviolet asymptotics of the Casimir scaling functions was calculated by perturbative solution of these equations, and 
by means of the perturbed conformal field-theory  technique, and compared with numerical results.
\end{abstract}

\section{Introduction}
The ground state energy $E(L)$ of the  (1+1)-dimensional massive relativistic quantum field theory (QFT) defined 
in the circle $\mathbb{R}/L$
 can be represented 
as:
\begin{equation}
E(L)= L \,{\mathcal E}_b + \frac{Y(u)}{L}.
\end{equation}
Here ${\mathcal E}_b$ is the bulk energy density, $L$ is the circle circumference, and $Y(u)$ is the {\it Casimir  finite-size  scaling function} depending on the 
scaling parameter $u=m L$, where $m$ is the mass of the lightest particle in the theory. 
This well-known fact (see e.g. \cite{Mussardo10} and references therein) follows immediately from  simple dimensional arguments. It is well known also \cite{Al_Z90}, that the  free energy per unit length $f(T)$ at a non-zero temperature $T>0$ of the same QFT defined in the infinite line $\mathbb{R}$ can be also expressed in terms of the Casimir scaling function: 
\begin{equation}\label{ft}
f(T) ={\mathcal E}_b + T^2 \, Y(u),
\end{equation}
with the  different meaning of the scaling parameter $u=m/T$, however.

The ground- and excited-state finite-size   spectra  in   (1+1)-dimensional
quantum field theories attract much interest in  recent decades. This interest stems from several reasons. 
First, studying of such spectra  gives a deep insight into the structure of  integrable QFT, yielding information about their 
monodromy properties \cite{BLZ96}, integrals of motions \cite{BLZ97}, and the   renormalization group 
flow \cite{AlZ_06}. Second, investigation of  the finite-size scaling is  crucial for  correct 
interpretation of the results of computer simulations, which are typically performed on finite-size systems. And third, the 
relativistic (1+1)-dimensional QFT can describe the dynamic and thermodynamic  properties of the quazi-one-dimensional 
magnetic crystals in the scaling region near their continuous quantum phase transition points.  Accordingly, the 
appropriate universal Casimir scaling function $Y(u)$ should determine due to \eref{ft} the free-energy temperature dependence  of such a crystal in the 
scaling regime \cite{Rut20}.

Different methods have been used to study the finite-size  scaling functions, such as the conformal perturbation theory \cite{Cardy86,Affleck_1989,Luk98}, and the truncated conformal space approach \cite{Tak98,Tak98PL}. 
{\color{black}{A rather effective approach to the finite-size scaling problem in integrable 
quantum spin chains and QFTs was introduced in the early 90-th by Batchelor, 
Kl\"umper, and Pearce \cite{KB90,Klum91},  and later used by Destri and de Vega for the sine-Gordon model.}}
In the pioneered works \cite{Des92,De94}, {\color{black}Destri and de Vega}  analysed 
the  inhomogeneous light-cone version of the six-vertex model, and derived in the continuous limit  the  non-linear integral equation  describing the ground-state energy of the sine-Gordon finite circle. Its Euclidean dynamics evolves in the 
cylinder having one spatial, and one (Euclidean) time direction. An alternative viewpoint \cite{De94,Al_Z90} on such a quantum dynamics is possible, in which one treats  the compact direction on the cylinder as the Matsubara direction. In this alternative picture, the nonlinear integral equations are interpreted as the Thermodynamic Bethe Ansatz (TBA) equations,
which determine the free energy of the infinite system at a non-zero temperature. 
Note, that similar nonlinear integral TBA equations {\color{black} {for different spin-chain models}}
 were obtained earlier by Kl\"umper, Batchelor, and Pearce \cite{Klum91,Kl92}. The approach  based  on the  nonlinear integral equations was later generalised and applied to describe   the finite-size  excited-energy spectra in integrable QFT.  
The latter subject attracted much attention and  has been studied by many authors \cite{Tak98,Tak98PL,Fio97,Des97,Tak99,Zinn_Justin_98,Tak2000,Heg20}. 

The important advantage of the  {\color{black} Destri-de Vega} (DDV) nonlinear integral equations 
 {\color{black}  for the sine-Gordon model} is that they  not only make possible rather accurate numerical calculations of  the finite-size energy spectra \cite{Fio97}, but also can be used to derive perturbatively the asymptotic expansions for these spectra in different limiting cases. For the ground-state energy $E(L)$ 
of the 
sine-Gordon circle of length $L$,  two such expansions at $L\to\infty$,  and at $L\to0$ were first studied by Destri and de Vega in \cite{De94}. Their perturbative analysis in the infrared (recursive) limit $L\to\infty$ was rather straightforward 
and will not be discussed here. The asymptotical analysis of the DDV equations in the opposite ultraviolet  (conformal) limit $L\to0$, which  is much more difficult, was presented  in Section 7 of \cite{De94}. The final result 
(1.13) of this Section for  the asymptotic expansion of the ground state energy $E(L)$ in this limit can be written as:
\begin{equation}\label{EL}
E(L)\,\mathop{=}\limits_{L\to0} \, L \,{\mathcal E}_b -\frac{\pi}{6 L}-\frac{m^2 L}{4}\,\cot\frac{\pi^2}{2\gamma}+\frac{ 1}{L} \,\sum_{n=0}^\infty C_n(\gamma)\, (m L)^{4 n\gamma/\pi}. 
\end{equation}
Here $m$ is the  soliton mass,  $\gamma\in(0,\pi) $ is the real parameter simply related with the coupling constant 
$\beta$, 
see equation \eref{gam} below.
The "repulsive" and "attractive" regimes of the sine-Gordon model are realized at $0<\gamma<\pi/2$, and at 
 $\pi/2<\gamma<\pi$, respectively.  
The bulk energy density ${\mathcal E}_b$ 
 in the first term in the right-hand side contains the non-universal 
 ultraviolet-divergent contribution. The second term describes the leading finite-size correction, 
 which agrees with the  Conformal Field Theory (CFT) prediction  for the case of the  central charge $c=1$.  The third and fourth terms
 in the right-hand side of  \eref{EL} represent the sub-leading finite-size corrections. 
 
 It turns out, however, that the original derivation of the expansion \eref{EL} described in Section 7 of \cite{De94} was
 not completely consistent being essentially based on a mistaken assumption. In the present work we reconsider derivation
 of  expansion \eref{EL} for the ground state energy in the ultraviolet  limit $L\to0$ in order to fill this gap.  
 Combing asymptotical analysis of the nonlinear TBA equations with the  perturbative CFT calculation, we confirm, that expansion \eref{EL}
  holds both in the attractive and repulsive regimes at all generic  $\gamma\in(0,\pi)$,
 apart from the points $\gamma_l=\frac{\pi}{2l}$, $l=1,2,\ldots$.    
 At these exceptional points, the diverging factor $\cot\frac{\pi^2}{2\gamma}$
 in the right-hand side of  \eref{EL}  should be replaced by the factor $\frac{2}{\pi l}  \ln u$. It is shown, that the sum in the right-hand side of 
 \eref{EL} starts in fact from the  $n=1$ term, and for the coefficient $C_1(\gamma)$ we obtained the 
 following explicit expression:
\begin{equation}\label{C1c0}
C_1(\gamma)=-\frac{(16\pi)^{1-2\gamma/\pi}}{4} \left[\frac{\Gamma\left(\frac{\pi}{2\gamma}\right)}{\Gamma\left(\frac{\pi}{2\gamma}-\frac{1}{2}\right)}
\right]^{4\gamma/\pi}
\left[\frac{\Gamma\left(1-\frac{\gamma}{\pi}\right)}{\Gamma\left(\frac{\gamma}{\pi}\right)}
\right]^{4}
\frac{\Gamma\left(\frac{2\gamma}{\pi}-1\right)}{\Gamma\left(2-\frac{2\gamma}{\pi}\right)}.
\end{equation}
This formula represents the main result of the present work.
Note, that the finite-size correction term $L^{-1}C_1(\gamma) (mL)^{4\gamma/\pi}$  dominates over the  term 
$-\frac{m^2 L }{4}\,\cot\frac{\pi^2}{2\gamma}$ in the repulsive regime $0<\gamma<\pi/2$ at $mL\to0$, while 
the cotangent term excesses the  infinite sum in the right-hand side of \eref{EL}
 in the attractive  case $\pi/2<\gamma<\pi$.

The rest of the paper is organised as follows. In the next Section, we recall few necessary facts
about the quantum sine-Gordon model. Section \ref{Sect3} is  addressed to  the sine-Gordon model
in the cylindric geometry. We describe there two versions of the nonlinear integral equations derived by Destri and de Vega 
\cite{De94},  then critically review their asymptotical analysis of the second (TBA) version of these equations in the ultraviolet
limit. It is shown, that this analysis in \cite{De94} was inconsistent and requires revision. The improved perturbative calculations 
of the Casimir scaling function $Y(u|\gamma)$ in the ultraviolet limit $u\to0$ are presented in Section \ref{UV}. 
In Section \ref{Num}, we describe the results of numerical calculations of the Casimir scaling function $Y(u|\gamma)$ obtained
by the  iterative solution of the DDV integral equations at different values of the parameter $\gamma$. 
At   small $u\ll1$,  these numerical results display a nice agreement with the analytical asymptotic dependencies
obtained in Section \ref{UV}. 
 Finally, there are three Appendices. In  Appendices A and B  we describe two alternative analytical calculations of the Casimir scaling function 
 $Y(u|\gamma)$  in the ultraviolet limit $u\ll1$. In  \ref{PTBA} we exploit to this end the small-$u$ asymptotical analysis of the nonlinear TBA equations, while in \ref{CFT} we use the perturbative CFT technique. In \ref{BLZ} 
 we recall some results of \cite{BLZ97} relating  to the "massless"case  $u=0$  of the DDV  equation, and 
 show that our findings are consistent with these results. 
\section{Quantum sine-Gordon model \label{QSG}}
The two-dimensional sine-Gordon  model can be  defined by the action \cite{Al_Z95},
\begin{equation}\label{HSG}
\mathcal{A}_{SG}=\frac{1}{16\pi} \int (\partial_a \varphi)^2\, d^2 x -2 \mu \int \cos(\beta \varphi)\, d^2 x,
\end{equation}
where $\varphi(x)$ is the scalar field in the two-dimensional Euclidean space-time with the coordinates $x^a=(x^0,x^1)$, 
and $0<\beta<1$ is the real parameter. We shall use also three related parameters $0<\gamma_s<\infty$, 
$0<\gamma<\pi$, and $0<\xi<\infty$,
\begin{eqnarray}\label{gam}
\gamma_s=\frac{8\pi \beta^2}{1-{\beta^2}},\quad\gamma=\pi\left(1-{\beta^2}\right),\quad
\xi=\frac{\beta^2}{1-\beta^2}.
\end{eqnarray}
The sine-Gordon model can be viewed as the perturbation of the Gaussian field theory by the 
exponential operators $\exp(\pm i \beta \varphi)$
having the scaling dimension 
\begin{equation}\label{scd}
{\mathrm x}=2{\beta^2}=\frac{2 \xi}{\xi+1}=\frac{2(\pi-\gamma)}{\pi}.
\end{equation}
These fields will be  normalised according to the conventional in the CFT condition \cite{Al_Z95},
\begin{equation}
\langle e^{ i \beta \varphi(x)} e^{-i \beta \varphi(0) }\rangle_{\mu=0}=
\frac{1}{|x|^{2{\mathrm x}}}.
\end{equation}

The particle content and the scattering properties of the sine-Gordon model are well known \cite{Zam77}.
In the repulsive regime at 
$\frac{1}{2}<\beta^2<1$, 
the model contains only the soliton and antisoliton excitations, which are the massive relativistic particles. 
Accordingly, their energy and momentum read,
\begin{equation}
\epsilon(\alpha)=m \cosh \alpha, \quad p(\alpha)=m \sinh \alpha,
\end{equation}
where $\alpha$ is the particle rapidity. 
In the attractive regime at $0<\beta^2<\frac{1}{2}$, the bound states of  solitons and antisolitons 
 also emerge in the theory.
 The dimensional coupling constant $\mu$ is related to the soliton mass $m$ as 
 follows:
 \begin{equation}\label{mum}
 \mu =\kappa(\xi) \,m^{2/(\xi+1)}. 
 \end{equation}
 The constant $\kappa(\xi)$ was found by Al.~B.~Zamolodchikov \cite{Al_Z95}:
  \begin{equation}
  \kappa(\xi)=\frac{1}{\pi}\frac{\Gamma\left(\frac{\xi}{\xi+1}\right)}{\Gamma\left(\frac{1}{\xi+1}\right)}
  \left[
  \frac{\sqrt{\pi}\,\Gamma\left(\frac{\xi+1}{2}\right)}{2\,\Gamma\left(\frac{\xi}{2}\right)}
  \right]^{2/(\xi+1)}.
   \end{equation}
The scattering matrix in the sine-Gordon model is known due to A.~B.~Zamolodchikov \cite{Zam77}. 
 In particular, the integral form of the soliton-soliton scattering amplitude reads \cite{Sm92},
\begin{equation}\label{Sc0}
S(\alpha, \gamma_s)=-\exp
\left[
-i\int_0^\infty \frac{dy}{ y} \frac{\sin(2 \alpha y)\sinh[(\pi-\frac	{\gamma_s}{8})y]}{\cosh (\pi y)\sinh(\gamma_s y/8)}\right].
 \end{equation}
 This scattering amplitude satisfies the following equality,
 \begin{equation}\label{Seq}
 S(\alpha, \gamma_s)\,S(\alpha+i \pi, \gamma_s)=-\frac{\sinh(8\pi \alpha/\gamma_s)}{\sinh[8\pi (\alpha+i\pi)/\gamma_s]}.
 \end{equation}
 Note, that the XYZ spin-1/2 chain model defined by the Hamiltonian 
\begin{equation}\label{XYZH}
{H}=\frac{1}{2}\sum_{j}\!\!\left(J_x\,\sigma_j^x\sigma_{j+1}^x+J_y\,\sigma_j^y\sigma_{j+1}^y+
J_z\,\,
\sigma_j^z\sigma_{j+1}^z
\right),
\end{equation}
is known to become equivalent to the sine-Gordon model in the continuous limit 
\begin{equation}\label{scl}
0<|J_z|<J_x <J_y , \quad J_y-J_x\ll J_x, \quad  J_z=J_x \cos \gamma. 
\end{equation}
\section{Sine-Gordon model on the cylinder and the DDV equation\label{Sect3}}
Let us now turn to the sine-Gordon model defined on the torus $\mathcal{T}$,
\begin{equation}
\mathcal{T}=\{x\in{\mathcal T}|0<x^0<L', 0<x^1<L\}
\end{equation}
with
the periodical boundary conditions,  
\[
\varphi(x^0,x^1)=\varphi(x^0,x^1+L)=\varphi(x^0+L',x^1),
\]
and proceed to the limit $L'\to\infty$ corresponding to the cylindric geometry. The partition function 
\begin{equation}\label{pf}
Z(\mu,L,L')=\int D[\varphi] \, \exp(-{\mathcal A}_{G}[\varphi ]) \exp\left\{2\mu \int_{\mathcal T}d^2x 
\cos[\beta\,\varphi(x)]\right\}
\end{equation}
 can be 
written in this limit as
\begin{equation}\label{ZZ}
Z(\mu,L,L')=\exp[-L' \, E(\mu,L)+O(1)],
\end{equation}
where ${\mathcal A}_{G}[\varphi ]$ is the action of the Gaussian free-field theory,  and 
$E(\mu,L)$  is the ground-state energy of the sine-Gordon model in the circle of length $L$.
The latter can be represented as
\begin{equation}\label{EY}
E(\mu,L)=L \,{\mathcal E}_b +\frac{Y(u|\gamma)}{L},
\end{equation}
where ${\mathcal E}_b={\mathcal E}_b(m,\gamma)$ is the bulk energy density, 
and $Y(u|\gamma)$ is the Casimir scaling function, which  depends  on the scaling parameter 
$u=m L$.

The scaling function $Y(u|\gamma)$ has the following explicit representation
\begin{equation}\label{Cas1}
Y(u|\gamma)= -\frac{u}{\pi }\int_{-\infty}^{\infty} d\alpha \, \sinh \alpha \cdot 
{\rm Im}\ln\left[1+\mathfrak{f}(\alpha+i 0|u,\gamma)\right],
\end{equation}
where $\mathfrak{f}(\alpha|u,\gamma)$ is the solution of the  DDV nonlinear integral equation 
\begin{eqnarray}\label{lnf2}
&&-i \ln {\mathfrak{f}}(\alpha|u,\gamma)=u \sinh \alpha+\\\nonumber
&&2\int_{-\infty}^{\infty}
d\alpha' \, G(\alpha-\alpha'|\gamma) \,{\rm Im}\,\ln[1+\mathfrak{f}(\alpha'+i 0|u,\gamma)].
\end{eqnarray}
The integral kernel $G(\alpha|\gamma)$ in the right-hand side of \eref{lnf2} is related with the soliton-soliton scattering amplitude \eref{Sc0},
\begin{equation}
\fl G(\alpha| \gamma)=\frac{1}{2 \pi i}\frac{\partial \ln S(\alpha,\gamma_{s})}{\partial\alpha}\bigg|_{\gamma_s=8\pi\left(\frac{\pi}{\gamma}-1\right)}=
\frac{1}{4\pi}
\int_{-\infty}^\infty {dk}\, \frac{\cos(\alpha k) \sinh\left[\left(\frac{\pi^2 }{2\gamma}-\pi\right)k\right]}
{\cosh\left(\frac{\pi k}{2}\right)\sinh\left[\left(\frac{\pi^2 }{2\gamma}-\frac{\pi}{2}\right)k\right]}\label{Sc}.
\end{equation}
Note the equality following from \eref{Seq},
\begin{equation}\label{eqQ}
G(\alpha +i\pi|\gamma)=-G(\alpha| \gamma)-g(\alpha| \gamma),
\end{equation}
where
\begin{equation}
g(\alpha| \gamma)=\frac{i\gamma}{2\pi(\pi-\gamma)}
\left[
\coth\left(
\frac{\gamma\,\alpha }{\pi-\gamma}
\right)-
\coth\left(
\frac{\gamma\,(i\pi+\alpha)}{\pi-\gamma}
\right)
\right].
\end{equation}
The exact representation \eref{Cas1}-\eref{Sc} for the Casimir scaling function of the sine-Gordon model was obtained by Destri and de Vega
\cite{Des92,De94} by  analysis of the Bethe-Ansatz equations for the six-vertex model in the light-cone approach. Later  Fioravanti and  Rossi
\cite{Fioravanti_2005} derived the same representation for $Y(u|\gamma)$ from the Bethe-Ansatz solution of the XYZ 
spin chain in the scaling limit \eref{scl}.

The integral representation  \eref{Cas1}-\eref{Sc} for  the Casimir scaling function 
$Y(u|\gamma)$ applies to the sine-Gordon model in the  
complete range  $0<\gamma<\pi$ of the parameter $\gamma$. In the repulsive regime $0<\gamma<\pi/2$, it can be 
modified \cite{De94} by means of the analytical continuation to the form 
\begin{equation}\label{Ysc}
Y(u|\gamma) =
- \frac{u}{\pi} \int_{-\infty}^\infty {d\alpha}\,\cosh\alpha\,\,{\rm Re}\,  \ln[1+e^{-\varepsilon(\alpha|u,\gamma)}], 
\end{equation}
where the function $\varepsilon(\alpha|u,\gamma)=- \ln {\mathfrak{f}}(\alpha+i \pi/2|u,\gamma)$ ({the pseudoenergy}) solves the system of the TBA integral equations,
\begin{eqnarray}\label{tBA1}
&& \varepsilon(\alpha|u,\gamma)=u \cosh \alpha
- \int_{-\infty}^{\infty}
d\alpha' \, G(\alpha-\alpha'|\gamma) \ln[1+e^{-\varepsilon(\alpha'|u,\gamma)}]\\
 &&+\int_{-\infty}^{\infty}d\alpha' \, G(\alpha-\alpha'+i \pi - i 0|\gamma)  \ln\big[1+e^{-\bar{\varepsilon}(\alpha'|u,\gamma)}\big],
 \nonumber\\\nonumber
&&\bar{\varepsilon}(\alpha|u,\gamma)=u \cosh \alpha
- \int_{-\infty}^{\infty}
d\alpha' \, G(\alpha-\alpha'|\gamma) \ln\big[1+e^{-\bar{\varepsilon}(\alpha'|u,\gamma)}\big]\\
&&+ \int_{-\infty}^{\infty}d\alpha' \, G(\alpha-\alpha'-i \pi + i 0|\gamma)  \ln\big[1+e^{-{\varepsilon}(\alpha'|u,\gamma)}\big],\nonumber
\end{eqnarray}
 and $\bar{\varepsilon}(\alpha|u,\gamma)$ is the complex conjugate of ${\varepsilon}(\alpha|u,\gamma)$.
Exploiting equality \eref{eqQ},  these equations can be rewritten in the equivalent form, 
 \begin{eqnarray}\label{tBA3}
&& \varepsilon(\alpha|u,\gamma)=u \cosh \alpha
-2  \int_{-\infty}^{\infty}
d\alpha' \, G(\alpha-\alpha'|\gamma)\,{\rm Re}\, \ln[1+e^{-\varepsilon(\alpha'|u,\gamma)}]\\
 &&-\int_{-\infty}^{\infty}d\alpha' \, g(\alpha-\alpha'- i 0|\gamma)  \ln\big[1+e^{-\bar{\varepsilon}(\alpha'|u,\gamma)}\big],
 \nonumber\\\nonumber
&&\bar{\varepsilon}(\alpha|u,\gamma)=u \cosh \alpha
- 2\int_{-\infty}^{\infty}
d\alpha' \, G(\alpha-\alpha'|\gamma)\,{\rm Re}\, \ln\big[1+e^{-\bar{\varepsilon}(\alpha'|u,\gamma)}\big]\\
&&- \int_{-\infty}^{\infty}d\alpha' \, g(\alpha-\alpha'+ i 0|\gamma)  \ln\big[1+e^{-{\varepsilon}(\alpha'|u,\gamma)}\big].\nonumber
\end{eqnarray}

In Section 7.3 of their article \cite{De94},  Destri and de Vega performed   the asymptotical analysis of the nonlinear integral equations \eref{tBA1} 
in order to describe the behaviour of the scaling function $Y(u| \gamma)$ at large and small $u$. 
Since equations \eref{tBA1} can be used only in the repulsive regime $0<\gamma<\pi/2$, their analysis could be related 
to this regime only. In what follows, we shall recall the main steps of this analysis in the ultraviolet limit $u\to0$.

Destri and de Vega wrote the solution of equations \eref{tBA1} at $u\ll1$ in the form,
\begin{equation}\label{ve}
{\varepsilon}(\alpha|u,\gamma)={\varepsilon}_k(\alpha-R(u)|\gamma)+{\varepsilon}_k(-\alpha-R(u)|\gamma)+
\eta{(\alpha|u,\gamma)},
\end{equation}
where $R(u)=\ln(2/u)$, and  the function 
\begin{equation}
{\varepsilon}_k(\alpha|\gamma)=\lim_{u\to0} {\varepsilon}(\alpha-R(u)|u,\gamma)
\end{equation} 
solves the integral equation
\begin{eqnarray}\label{tBAk}
\varepsilon_k(\alpha|\gamma)=e^\alpha
- \int_{-\infty}^{\infty}
d\alpha' \, G(\alpha-\alpha'|\gamma) \ln[1+e^{-\varepsilon_k(\alpha'|\gamma)}]
 +\\\nonumber
 \int_{-\infty}^{\infty}d\alpha' \, G(\alpha-\alpha'+i \pi - i 0|\gamma)  \ln\big[1+e^{-\bar{\varepsilon}_k(\alpha'|\gamma)}\big].
\end{eqnarray}
The function $\eta{(\alpha|u,\gamma)}$ was treated as a small correcting term vanishing at $u\to 0$.

A similar representation was used in \cite{De94} for the logarithm function
\begin{equation}\label{L}
L(\alpha|u,\gamma)=L_k(\alpha-R(u)|\gamma)+L_k(-\alpha-R(u)|\gamma)-\ln 2+l{(\alpha|u,\gamma)},
\end{equation} 
where
\begin{equation}\label{Llog}
L(\alpha|u,\gamma)=\ln[1+e^{-\varepsilon(\alpha|u,\gamma)}], \quad
L_k(\alpha|\gamma)=\ln[1+e^{-\varepsilon_k(\alpha|\gamma)}],
\end{equation}
and $l{(\alpha|u,\gamma)}$ was also supposed to vanish at $u\to 0$.
Substitution of \eref{L} into \eref{Ysc} leads after some manipulations to the following representation of the 
Casimir scaling function:
\begin{eqnarray}\label{dY}
&&Y(u|\gamma)=-\frac{1}{2\pi}\int_{-\infty}^\infty d\alpha \,e^\alpha \,{\mathrm {Re}}\,L_k(\alpha|\gamma)
-\\
&&\frac{u^2}{2\pi}\int_{-\infty}^\infty d\alpha \,e^{-\alpha} \,{\mathrm {Re}}\,\partial_\alpha L_k(\alpha|\gamma)-
\frac{u}{\pi}\int_{-\infty}^\infty d\alpha\, \cosh \alpha  \,{\mathrm {Re}}\,l{(\alpha|u,\gamma)}.\nonumber
\end{eqnarray}
The first term in the right-hand side calculated by means of the well-known dilogarithm-function trick 
gives the CFT-predicted value, 
\[
-\frac{1}{2\pi}\int_{-\infty}^\infty d\alpha \,e^\alpha \,{\mathrm {Re}}\,L_k(\alpha|\gamma)=-\frac{\pi}{6}.
\]
In order to compute the second term in the right-hand side of \eref{dY}, the following arguments were
adopted. To provide convergency of the second integral  
in the right-hand side of \eref{dY}
at $\alpha\to-\infty$, the function $\varepsilon_k(\alpha|\gamma)$ must decay faster than $e^\alpha$.
After differentiation of equation \eref{tBAk}, one obtains:
\begin{eqnarray}\label{dk}
\partial_\alpha\, \varepsilon_k(\alpha|\gamma)=e^\alpha
- \int_{-\infty}^{\infty}
d\alpha' \, G(\alpha-\alpha'|\gamma)\, \partial_{\alpha'}\,L_k(\alpha'|\gamma)
 +\\\nonumber
 \int_{-\infty}^{\infty}d\alpha' \, G(\alpha-\alpha'+i \pi - i 0|\gamma) \, \partial_{\alpha'}\,\overline{L_k(\alpha'|\gamma)}.
\end{eqnarray}
Then adopting the asymptotic expansion of the kernel 
$G(\alpha|\gamma)$ at large $|\alpha|$
\begin{equation}
G(\alpha|\gamma)=\mathfrak{a}_1\,e^{-|\alpha|}+\mathfrak{b}_1e^{-2\hat{\gamma}|\alpha|/\pi}+\ldots,
\end{equation}
where $\hat{\gamma}=\gamma (1-\gamma/\pi)^{-1}$ and 
\[
\mathfrak{a}_1=\frac{1}{\pi}\tan\left(\frac{\pi^2}{2\gamma}\right), \quad 
\mathfrak{b}_1=\frac{\hat{\gamma}}{\pi^2}\tan\left(\frac{\pi^2}{\pi-\gamma}\right),
\]
 and proceeding to the limit $\alpha\to-\infty$ in equation  \eref{dk}, Destri and de Vega obtained
\begin{equation}
\fl \partial_\alpha\, \varepsilon_k(\alpha|\gamma)\simeq
\left[
1-2 \mathfrak{a}_1 \int_{-\infty}^\infty d\alpha'\, e^{-\alpha'}  \,{\mathrm {Re}}\,\partial_{\alpha'}\,L_k(\alpha'|\gamma)
\right]\, e^{\alpha}+O(e^{2\alpha})+O(e^{2\hat{\gamma}\alpha/\pi}).
\end{equation}
Putting the coefficient of $(e^\alpha)$ to zero, they conclude that
\begin{equation}
 \int_{-\infty}^\infty d\alpha'\, e^{-\alpha'}  \,{\mathrm {Re}}\,\partial_{\alpha'}\,L_k(\alpha'|\gamma)=\frac{1}{2\mathfrak{a}_1}, 
\end{equation}
and
\begin{equation}\label{YDdV}
Y(u|\gamma)=-\frac{\pi}{6}-\frac{u^2}{4}\, \cot\frac{\pi^2}{2\gamma}-\frac{u}{\pi}\int_{-\infty}^\infty d\alpha\, \cosh \alpha  \,{\mathrm {Re}}\,l{(\alpha|u,\gamma)}.
\end{equation}

The small-$u$ asymptotics of the third term in the right-hand side of \eref{YDdV} was studied in 
Section 7.4 of  \cite{De94}. Destri and de Vega
came there to the conclusion, that this term admits at $u\to0$ the asymptotical expansion of the form 
\begin{equation}
-\frac{u}{\pi}\int_{-\infty}^\infty d\alpha\, \cosh \alpha  \,{\mathrm {Re}}\,l{(\alpha|u,\gamma)}=
\sum_{n=0}^\infty C_n(\gamma)\, u^{4 n\gamma/\pi}.
\end{equation}
So, the final result of \cite{De94} for the ultraviolet asymptotics $u\to0$ of the Casimir scaling function in the sine-Gordon model in the repulsive regime takes the form:
\begin{equation}\label{YDdV1}
Y(u|\gamma)=-\frac{\pi}{6}-\frac{u^2}{4}\, \cot\frac{\pi^2}{2\gamma}+\sum_{n=0}^\infty C_n(\gamma)\, u^{4 n\gamma/\pi}.
\end{equation}
No explicit  expressions for the coefficients $C_n(\gamma)$  were given in \cite{De94}.

It turns out, however,  that  the derivation of \eref{YDdV1} given in \cite{De94} and sketched above
is inconsistent and contains several mistakes. 
First, equation \eref{ve} representing the  pseudoenergy ${\varepsilon}(\alpha|u,\gamma)$ can be useful in the asymptotical analysis, if  two initial terms in its right-hand side 
dominate in the limit $u\to0$, while the term $\eta(\alpha|u,\gamma)$ describes the small correction. However, this is
not the case, since the pseudoenergy $\varepsilon(\alpha|u,\gamma)$ is the complex-valued function having the reflection symmetry
\begin{equation}\label{refl}
\varepsilon(\alpha|u,\gamma)=\overline{\varepsilon(-\alpha|u,\gamma)},
\end{equation}
which is not respected by the function ${\varepsilon}_k(\alpha-R(u)|\gamma)+{\varepsilon}_k(-\alpha-R(u)|\gamma)$ in the 
right-hand side of  \eref{ve}. So, instead of  \eref{ve}, one should write
\begin{equation}\label{ve1}
{\varepsilon}(\alpha|u,\gamma)={\varepsilon}_k(\alpha-R(u)|\gamma)+\bar{{\varepsilon}}_k(-\alpha-R(u)|\gamma)+
\eta{(\alpha|u,\gamma)}.
\end{equation}
Equation \eref{L} for the logarithm functions requires the same correction:
\begin{equation}\label{L1}
L(\alpha|u,\gamma)=L_k(\alpha-R(u)|\gamma)+\bar{L}_k(-\alpha-R(u)|\gamma)-\ln 2+l{(\alpha|u,\gamma)},
\end{equation} 
where $\bar{L}_k(-\alpha-R(u)|\gamma)=\overline{{L}_k(-\alpha-R(u)|\gamma)}$.

There is, however, one more problem with equation  \eref{L}, which remains also in its improved version
 \eref{L1}. Its zero-order part
\begin{equation}
 L^{(0)}(\alpha|u,\gamma)=L_k(\alpha-R(u)|\gamma)+\bar{L}_k(-\alpha-R(u)|\gamma)-\ln 2
\end{equation}
 approximates well at small $u$ the exact function  $L(\alpha|u,\gamma)$ in the interval $-R(u)<\alpha<R(u)$, but is not 
 appropriate at larger $|\alpha|>R(u)$. Really, one can easily see that the pseudoenergy $\varepsilon(\alpha|u,\gamma)$ exponentially 
 increases at large $|\alpha|\to\infty$, namely
 \[
 \varepsilon(\alpha|u,\gamma)\simeq e^{|\alpha|-R(u)}\quad {\rm {at}}\; |\alpha|-R(u)\gg1.
\]
Accordingly, the difference $L(\alpha|u,\gamma)-\ln 2$ decays extremely  fast at large  $|\alpha|$, as
\begin{equation}\label{LL}
L(\alpha|u,\gamma)-\ln 2=-\frac{1}{2}\, \exp[-e^{|\alpha|-R(u)}]+\ldots,\quad {\rm {at}}\; |\alpha|-R(u)\gg1.
\end{equation}
In contrast, the difference $L^{(0)}(\alpha|u,\gamma)-\ln 2$ decays much slower at $|\alpha|\to\infty$.
Really, if we assume that ${\varepsilon}_k(\alpha|\gamma)$ exponentially decays at large negative $\alpha$,
\begin{equation}\label{de}
{\varepsilon}_k(\alpha|\gamma)\simeq \frac{b(\gamma)}{a(\gamma)}\,e^{a(\gamma) \alpha } \quad {\rm {at}}\; \alpha\to-\infty,
\end{equation}
with some  $a(\gamma)>0$, we get
\[
L^{(0)}(\alpha|u,\gamma)-\ln 2=-\frac{\overline{b(\gamma)} }{2a(\gamma)}\,e^{-a(\gamma)\,[\alpha+R(u)]}+\ldots,\quad {\rm {at}}\; \alpha-R(u)\gg1,
\]
instead of \eref{LL}.

We shall see in the next Section that  assumption \eref{de} is indeed correct, and that 
\begin{equation}\label{a}
a(\gamma)=\frac{2\gamma}{\pi}.
\end{equation} 
Therefore, the integrand in the second integral in the right-hand side of \eref{dY} behaves 
at $\alpha\to-\infty$ as
\begin{equation}\label{ig}
\,e^{-\alpha} \,{\mathrm {Re}}\,\partial_\alpha L_k(\alpha|\gamma)\simeq-\frac{ \,{\mathrm {Re}} \,b(\gamma)}{2}\, \exp\left[
\left(\frac{2 \gamma}{\pi}-1\right)\alpha
\right].
\end{equation}
Since $\left(\frac{2 \gamma}{\pi}-1\right)<0$ in the repulsive case $0<\gamma<\pi/2$, the integrand \eref{ig} exponentially
increases at $\alpha\to-\infty$, and  the second integral in the right-hand side of \eref{dY} {\it {diverges}} at $\alpha\to-\infty$. 
Therefore, the very assumptions  of Destri and de Vega  about convergency of that integral is not valid, and 
 their subsequent analysis based on this assumption is not satisfactory.
 
 In the next Section, we present the consistent derivation of the small-$u$ asymptotic expansion for the 
 Casimir scaling function $Y(u|\gamma)$, which is free from the problems outlined above.
\section{Ultraviolet asymptotics of the Casimir scaling function \label{UV}}
Before proceeding to the calculations, let us summarise the obtained results.
For generic values of $\gamma\in(0,\pi)$, the Casimir scaling function $Y(u|\gamma)$ has the small-$u$ asymptotical expansion of the form \eref{YDdV1} predicted by 
Destri and de Vega, with $C_0(\gamma)\equiv 0$, and $C_1(\gamma)$ is given by equation \eref{C1c0}.
The asymptotic formula
\begin{equation}\label{Y}
Y(u|\gamma)+\frac{\pi}{6}=C_1(\gamma)\, u^{4\gamma/\pi}-\frac{u^2}{4} \cot\left(\frac{\pi^2}{2 \gamma}\right)+\sum_{n=2}^\infty C_n(\gamma)\, u^{4n \gamma/\pi}
\end{equation}
holds at $u\to 0$ 
both in the repulsive and attractive cases of the 
sine-Gordon model for all values of the parameter $\gamma$ in the interval $0<\gamma<\pi$, apart from the points $\gamma_l=\frac{\pi}{2 l}$, 
with $l=1,2,\ldots$.

The first term in the right-hand side of \eref{Y} dominates over the second one in the repulsive regime $0<\gamma<\pi/2$. In the 
attractive regime $\pi/2<\gamma<\pi$, the second term becomes larger than the first one.
\begin{figure}
\includegraphics[width=1.\linewidth]{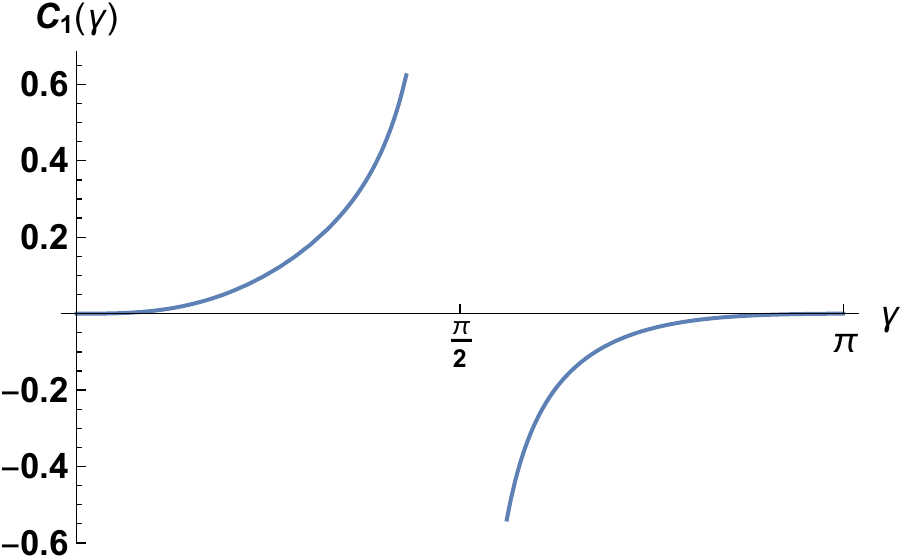}
\caption{Plot of the coefficient $C_1(\gamma)$ defined by \eref{C1c0}.}
\label{C1pl}
\end{figure}
Plot of the coefficient $C_1(\gamma)$ defined by \eref{C1c0} is shown in Figure \ref{C1pl}. This 
function has a simple pole in the free-fermionic point $\gamma=\pi/2$. At this point, formula \eref{Y}
should be replaced by 
\begin{equation}\label{YFF}
Y(u|\pi/2)+\frac{\pi}{6}=\frac{u^2}{2\pi}
\left(
-\ln \frac{u}{\pi}+\frac{1}{2}-\gamma_E
\right)+O(u^4),
\end{equation}
where $\gamma_E$ is the Euler's constant.
At the points $\gamma_l=\frac{\pi}{2 l}$ 
with $l=2,3,\ldots$, equation \eref{Y} is to be replaced by
\begin{equation}\label{Y3}
Y(u|\gamma_l)+\frac{\pi}{6}=C_1(\gamma_l)\, u^{2/l}-\frac{u^2}{2\pi l}[\ln u+O(1)]+O(u^{4/l}).
\end{equation}

Proceeding to derivation of these results, we  start from the conformal limit $u=0$ in the repulsive case $0<\gamma<\pi/2$, and rewrite 
equation \eref{dk} in the form
\begin{eqnarray}\label{dkA}
\partial_\alpha\, \varepsilon_k(\alpha|\gamma)=e^\alpha
-2 \int_{-\infty}^{\infty}
d\alpha' \, G(\alpha-\alpha'|\gamma)\,{\mathrm{Re}}\, \partial_{\alpha'}\,L_k(\alpha'|\gamma)
 -\\\nonumber
 \int_{-\infty}^{\infty}d\alpha' \, {g}(\alpha-\alpha'- i 0|\gamma) \, \partial_{\alpha'}\,\overline{L_k(\alpha'|\gamma)},
\end{eqnarray}
using equality \eref{eqQ}.
We  suppose, that the pseudoenergy $\varepsilon_k(\alpha|\gamma)$ corresponding to the critical point $u=0$ exponentially 
decays at
large negative $\alpha$ according to \eref{de}, with some positive exponent $0<a(\gamma)<1$. 
Then we proceed to the limit of a large negative $\alpha$, replace the derivative of the logarithm function $L_k(\alpha'|\gamma)$ 
defined by \eref{Llog} by the 
first term in its Taylor expansion,
\[
\partial_{\alpha'} L_k(\alpha'|\gamma)=-\frac{ \partial_{\alpha'}\varepsilon_k(\alpha'|\gamma)}{2}+\ldots
\]
and neglect the term $e^\alpha$ in the right-hand-side of equation \eref{dkA}. As the result, we obtain 
the linear uniform integral equation
\begin{eqnarray}\label{dkB}
\partial_\alpha\, \varepsilon_k(\alpha|\gamma)=
 \int_{-\infty}^{\infty}
d\alpha' \, G(\alpha-\alpha'|\gamma)\,{\mathrm{Re}}\, \partial_{\alpha'}\,\varepsilon_k(\alpha'|\gamma)
 +\\\nonumber
\frac{1}{2} \int_{-\infty}^{\infty}d\alpha' \, {g}(\alpha-\alpha'- i 0|\gamma) \, \partial_{\alpha'}\,\overline{\varepsilon_k(\alpha'|\gamma)},
\end{eqnarray}
which determines the asymptotics of the pseudoenergy $\varepsilon_k(\alpha|\gamma)$ at $\alpha\to-\infty$, 
up to a real numerical factor. 
The solution of this equation indeed has the form consistent with \eref{de}, namely
\numparts
\begin{eqnarray}\label{deps}
&\partial_\alpha\varepsilon_k(\alpha|\gamma)=b\, e^{2\alpha\gamma/\pi},\\
&b=- i \,c_1(\gamma) \, e^{i\gamma}.\label{bb}
\end{eqnarray}
\endnumparts
The real numerical factor $c_1(\gamma)$ is in fact positive. It will be   determined later, see equation \eref{eq:c1}.  
In deriving \eref{deps}, \eref{bb} we have used two explicit integral formulas, which are valid at $0<\gamma<\pi/2$ and 
$0<a<\min(1,\hat{\gamma})$, 
\numparts\label{equ}
\begin{eqnarray}
&\int_{-\infty}^\infty d\alpha\, e^{a\alpha}\,g(-\alpha- i 0|\gamma)=-e^{\pi i a/2}\,\frac{\sin[\pi a({\nu}^{-1}-2^{-1})]}{\sin(\pi a/ {\nu})},\\
&\int_{-\infty}^\infty d\alpha\, e^{a\alpha}\,G(-\alpha|\gamma)=\frac{\sin[\pi a({\nu}^{-1}-2^{-1})]}{2\cos(\pi a/2)\sin(\pi a/ {\nu})},\label{argb}
\end{eqnarray}
\endnumparts
where $\nu=\frac{2\gamma}{\pi-\gamma}$.

Let us turn now to  the asymptotical behaviour of the pseudoenergy function  ${\varepsilon}(\alpha|u,\gamma)$ at a small nonzero scaling parameter $u\ll1$. 
In this limit, one can neglect to the leading order in $u$ the term $\eta{(\alpha|u,\gamma)}$ in 
equation \eref{ve1} and represent this function  as
\begin{equation}\label{ve3}
{\varepsilon}(\alpha|u,\gamma)\approx {\varepsilon}_k(\alpha-R(u)|\gamma)+\bar{{\varepsilon}}_k(-\alpha-R(u)|\gamma).
\end{equation}
At $|\alpha|\ll R(u)$, we can use for the function ${\varepsilon}_k(\alpha|\gamma)$ the asymptotical formula \eref{de}, 
and reduce  \eref{ve3} to the form
\begin{equation}\label{ve4}
{\varepsilon}(\alpha|u,\gamma)\approx \left(\frac{u}{2}\right)^{a(\gamma)}\, \frac{b\, e^{a(\gamma)\alpha}+\bar{b}\,e^{-a(\gamma)\alpha}}{a(\gamma)}, 
\end{equation}
with $a(\gamma)$ given by \eref{a}. 
Combining this with \eref{bb}, one finds,
\begin{eqnarray}\label{RI}
{\mathrm{Re}}\,{\varepsilon}(\alpha|u,\gamma)\approx   \left(\frac{u}{2}\right)^{2\gamma/\pi}\,\frac{2 c_1(\gamma) \cosh (2\gamma \alpha/\pi)}{a(\gamma)}\, \sin\gamma,\\
{\mathrm{Im}}\,{\varepsilon}(\alpha|u,\gamma)\approx  - \left(\frac{u}{2}\right)^{2\gamma/\pi}\,\frac{2 c_1(\gamma) \sinh(2\gamma \alpha/\pi)}{a(\gamma)}\, \cos\gamma.\nonumber
\end{eqnarray}
Thus, the ratios ${\mathrm{Re}}\,{\varepsilon}(\alpha|u,\gamma)/\cosh (2\gamma \alpha/\pi)$ and ${\mathrm{Im}}\,{\varepsilon}(\alpha|u,\gamma)/\sinh (2\gamma \alpha/\pi)$ must have at $u\ll1$ wide plateaus near the origin $\alpha=0$ at the values 
\begin{eqnarray}\label{ReE}
\frac{{\mathrm{Re}}\,{\varepsilon}(0|u,\gamma)}{\cosh (2\gamma \alpha/\pi)}\approx 
\left(\frac{u}{2}\right)^{2\gamma/\pi}\,\frac{\pi c_1(\gamma) }{\gamma}\, \sin\gamma,\\\nonumber
\frac{{\mathrm{Im}}\,{\varepsilon}(0|u,\gamma)}{\sinh (2\gamma \alpha/\pi)}\approx -
\left(\frac{u}{2}\right)^{2\gamma/\pi}\,\frac{\pi c_1(\gamma) }{\gamma}\, \cos\gamma
\end{eqnarray}
and 
\begin{equation}\label{rr}
\lim_{u\to 0}\lim_{\alpha\to 0}\frac{{\mathrm{Re}}\,{\varepsilon}(\alpha|u,\gamma)}{\cosh (2\gamma \alpha/\pi)}/
\frac{{\mathrm{Im}}\,{\varepsilon}(\alpha|u,\gamma)}{\sinh (2\gamma \alpha/\pi)}
=-\tan \gamma.
\end{equation}
These properties of the pseudoenergy ${\varepsilon}(\alpha|u,\gamma)$ at a small $u$ in the repulsive 
regime $0<\gamma<\pi/2$ were checked in  numerical calculations described in the next Section, see Figure
\ref{ratio}.

The outlined above perturbative analysis of the integral equation \eref{dkA} at $\alpha\to-\infty$ can be extended 
to higher orders in  $\exp(2 \alpha \gamma/\pi)$. It leads to the following asymptotic expansion for the pseudoenergy
$\varepsilon_k(\alpha|\gamma)$ at  $\alpha\to-\infty$:
\begin{equation}\label{vek}
\varepsilon_k(\alpha|\gamma)=-\frac{i \pi}{2 \gamma} \, \sum_{n=1}^\infty \frac{c_n(\gamma)}{n}  \exp\left[\frac{2 n \gamma}{\pi}\left(\alpha+\frac{i \pi}{2} \right)\right],
\end{equation}
with real coefficients $c_n(\gamma)$. Note, that the function 
\begin{equation}\label{fep}
{\mathfrak{f}}_k(\alpha|\gamma)=\exp[-\varepsilon_k(\alpha-i \pi/2|\gamma)]
\end{equation}
solves the "massless" version of the DDV equation:
\begin{eqnarray}\label{mles}
-i \ln {\mathfrak{f}}_k(\alpha|\gamma)=e^\alpha+
2\int_{-\infty}^{\infty}
d\alpha' \, G(\alpha-\alpha'|\gamma) \,{\rm Im}\,\ln[1+\mathfrak{f}_k(\alpha'+i 0|\gamma)].
\end{eqnarray}
The  "counting  function" $\phi_k(\alpha|\gamma)=-i \ln {\mathfrak{f}}_k(\alpha|\gamma)$ in the left-hand side of this equation 
is real at real $\alpha$. Its asymptotic expansion at $\alpha\to-\infty$ can be read from \eref{vek}, \eref{fep}:
\begin{equation}\label{vek1}
\phi_k(\alpha|\gamma)=\frac{ \pi}{2 \gamma} \, \sum_{n=1}^\infty \frac{c_n(\gamma)}{n}  e^{2 n \gamma \alpha/\pi}.
\end{equation}
It was shown in \cite{BLZ97}, that the same expansion for this counting function 
holds also in the attractive case $\pi/2<\gamma<\pi$, see the discussion 
below in  \ref{BLZ}.

\begin{figure}
\includegraphics[width=1.\linewidth]{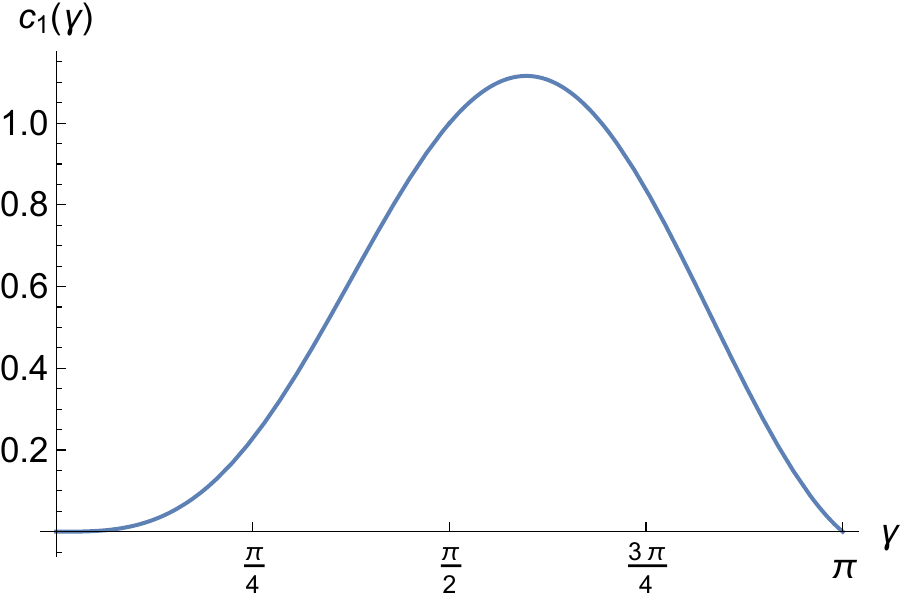}
\caption{The coefficient $c_1(\gamma)$ determined by equation \eref{eq:c1}.}
\label{c1a}
\end{figure}

Now let us return to the asymptotic expansion of the Casimir scaling functions $Y (u|\gamma)$
in the ultraviolet limit $u\to0$.  By means of the perturbative solution of the TBA equations \eref{tBA1}, we obtained 
at $0<\gamma<\pi/2$  the first sub-leading term in this expansion:
\begin{equation}\label{Yas}
Y(u|\gamma)+\frac{\pi}{6}=\frac{\pi^2 [c_1(\gamma)]^2}{4 \gamma^2 \sin(2\gamma)}\,(u/2)^{4\gamma/\pi}+\ldots.
\end{equation}
 Derivation of this asymptotic formula  is described in  \ref{PTBA}. 
The  coefficient $c_1(\gamma)$ in \eref{Yas} is the same as in expansions \eref{vek}, \eref{vek1}.
It turns out, that the direct  calculation of the constant $c_1(\gamma)$ from equation \eref{tBAk} presents a
difficult problem. By this reason, we 
determined the coefficient $C_1(\gamma)$ in a completely different way exploiting the perturbative CFT technique. 
This calculation is described in   \ref{CFT}, and the resulting expression for the coefficient $C_1(\gamma)$
is given in equation \eref{C1c0}. Using this result, one can recover\begin{footnote}{
In fact, equation \eref{Yas} determines $c_1(\gamma)$ up to the sign. The latter is fixed by the 
known value $c_1(\pi/2)=1$ at the free-fermionic point $\gamma=\pi/2$, where  $\phi_k(\alpha|\pi/2)=e^\alpha.$}
\end{footnote}  the coefficient $c_1(\gamma)$ 
 from equations \eref{Y},  \eref{Yas}:
\begin{eqnarray}\label{eq:c1}
c_1(\gamma)=4\pi^{3/2}\,\frac{\gamma \cot \gamma}{\pi-2 \gamma}\,\frac{\Gamma(\frac{1}{2}+\frac{\gamma}{\pi})}{[\Gamma(\frac{\gamma}{\pi})]^3} \left[
\frac{\Gamma(\frac{\pi}{2\gamma})}{\sqrt{\pi}\,\Gamma(-\frac{1}{2}+\frac{\pi}{2\gamma})
}
\right]^{2\gamma/\pi}=\\
4 \pi^{3/2}\,\left(\frac{2}{\pi}\right)^{2\gamma/\pi}\frac{\gamma\,\cos\gamma\,\, \Gamma\left(
\frac{1}{2}+\frac{\gamma}{\pi}\right)}{(\pi-2\gamma)\,\Gamma\left(
\frac{\gamma}{\pi}\right)}\, \kappa(\xi)\big|_{\xi=(\pi-\gamma)/\gamma}
.\nonumber
\end{eqnarray}
Plot of this function  is shown in Figure \ref{c1a}. 
The function $c_1(\gamma)$ is analytical in the whole interval $0<\gamma<\pi$, and has the essential 
singularity at $\gamma=0$. 

Though our derivation of formula \eref{eq:c1} for the coefficient $c_1(\gamma)$ was limited to the repulsive case $0<\gamma<\pi/2$, it remains valid
in the attractive regime $\pi/2<\gamma<\pi$ as well. In the latter case, the TBA equation in the form \eref{tBAk} does not 
hold any more, and one should define the coefficients $c_n(\gamma)$ from the expansion
\eref{vek1} for the solution of the DDV equation \eref{mles}. This integral equation was studied in the 
attractive  regime $\pi/2<\gamma<\pi$ to much details by 
Bazhanov, Lukyanov, and Zamolodchikov \cite{BLZ97}. 
It turns out, that the explicit expression for the coefficient $c_1(\gamma)$ in the attractive regime, which can be gained from \cite{BLZ97}, coincides exactly with our result \eref{eq:c1}, analytically continued into the interval  $\pi/2<\gamma<\pi$.
The details are given  in  \ref{BLZ}, were we recall some of results of the work \cite{BLZ97}.

\section{Numerical work \label{Num}}
The results described in the previous Section were confirmed by numerical calculations both
in the repulsive  and attractive regimes. 

In the repulsive regime $0<\gamma<\pi/2$, we calculated the pseudoenergy ${\varepsilon}(\alpha|u,\gamma)$
by iterative solution of the nonlinear TBA equations \eref{tBA1}, and then obtained the Casimir scaling function
$Y(u|\gamma)$ using the integral formula \eref{Ysc}.

In the attractive regime $\pi/2<\gamma<\pi$, the nonlinear integral equation  \eref{lnf2} is still valid, but the TBA equations in the form  \eref{tBA1} do not hold any more. The reason is that the integral kernel 
$G(\alpha| \gamma)$ defined by \eref{Sc} has the $t$-channel poles at $\alpha_n=in\pi(\frac{\pi}{\gamma}-1)$, with 
$n=1,2\ldots$.  In the attractive regime, one or several such poles come  
into the physical strip $0<{\mathrm {Im}}\, \alpha<\pi$ indicating appearance of the soliton-antisoliton 
bound states in the particle spectrum\begin{footnote}
{These bound states have the masses \cite{Zam77} $m_n=2 m \sin (\pi n \xi/2) =2 m\cosh[(i \pi -\alpha_n)/2]$.}
\end{footnote}. 
\begin{figure}
\includegraphics[width=1.\linewidth]{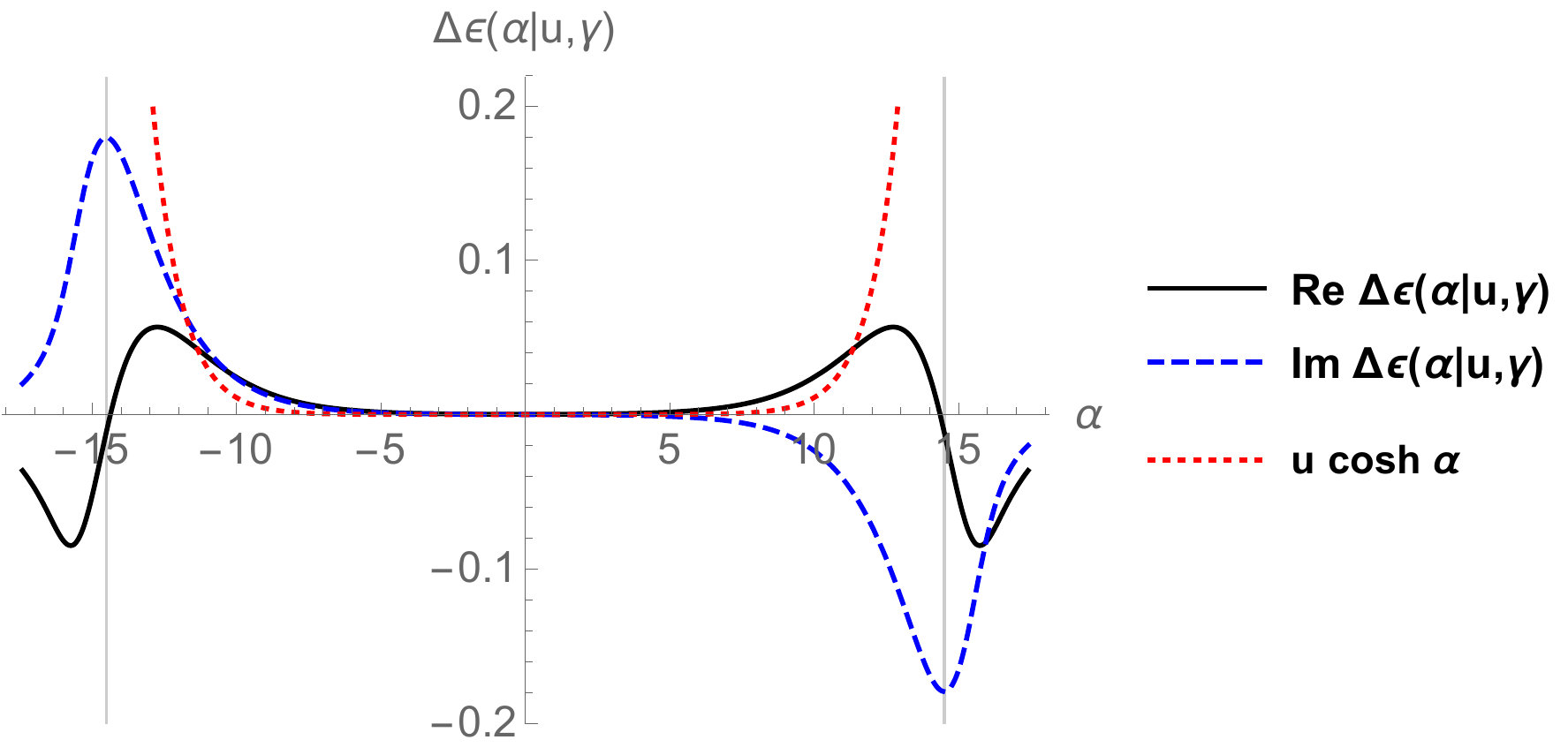}
\caption{Real (solid black)  and imaginary  (dashed blue) parts of the function $\Delta \varepsilon(\alpha|u,\gamma)$ 
defined by \eref{delte} versus rapidity $\alpha$ at $\gamma=0.3 \pi$ and  
$u=10^{-6}$.  Vertical lines are located at  
$\pm R(u)$, with $R(u)\approx 14.5.$}
\label{dev}
\end{figure}

These poles  prevent analytical 
continuation of equation \eref{lnf2} from the real $\alpha$-axis into the lines ${\mathrm {Im}}\, \alpha=\pm\pi/2$, 
which has been used in derivation of the TBA equation \eref{tBA1} in the repulsive case. To avoid this problem, 
we have used for numerical calculations in the attractive case the modified TBA equations, which were 
obtained from  \eref{lnf2} by analytical continuation in the rapidity variable $\alpha$ from 
the real axis into the line
${\mathrm {Im}}\, \alpha=\delta$, with some $\delta$  lying in the interval 
\begin{equation}\label{del}
\delta\in \left(0,\frac{\pi(\pi-\gamma)}{2\gamma}\right).
\end{equation}
These modified TBA equations read as,
\begin{eqnarray}\nonumber
& \varepsilon(\alpha|u,\gamma,\delta)=-i u \sinh (\alpha+i\delta)
- \int_{-\infty}^{\infty}
d\alpha' \, G(\alpha-\alpha'|\gamma) \ln[1+e^{-\varepsilon(\alpha'|u,\gamma,\delta)}]\\\label{tBA4}
 &+\int_{-\infty}^{\infty}d\alpha' \, G(\alpha-\alpha'+2i \delta)  \ln\big[1+e^{-\bar{\varepsilon}(\alpha'|u,\gamma,\delta)}\big],
\\\nonumber
&\bar{\varepsilon}(\alpha|u,\gamma,\delta)=i u \sinh (\alpha-i\delta)
- \int_{-\infty}^{\infty}
d\alpha' \, G(\alpha-\alpha'|\gamma) \ln\big[1+e^{-\bar{\varepsilon}(\alpha'|u,\gamma,\delta)}\big]\\
&+ \int_{-\infty}^{\infty}d\alpha' \, G(\alpha-\alpha'-2i \delta)  \ln\big[1+e^{-{\varepsilon}(\alpha'|u,\gamma,\delta)}\big],\nonumber
\end{eqnarray}
where 
\[
{\varepsilon}(\alpha|u,\gamma,\delta)=-\ln f(\alpha+i \delta|u,\gamma)
\]
is the modified pseudoenergy, 
and $\bar{\varepsilon}(\alpha|u,\gamma,\delta)$ is its complex conjugate,
\[
\bar{\varepsilon}(\alpha|u,\gamma,\delta)=-\overline{\ln f(\alpha+i \delta|u,\gamma)}=
\ln {f(\alpha-i \delta|u,\gamma)}.
\]
The Casimir scaling function $Y(u|\gamma)$ can be expressed in terms of the modified pseudoenergy
as follows,
\begin{equation}\label{YscA}
Y(u|\gamma) =
- \frac{u}{\pi}\, \,{\mathrm {Im}}\, \int_{-\infty}^\infty {d\alpha}\,\sinh(\alpha+i\delta)\, \ln[1+e^{-\varepsilon(\alpha|u,\gamma,\delta)}]. 
\end{equation}
The integral in the right-hand side  in fact  does not depend on the parameter $\delta$ provided 
the latter lies in the allowed interval \eref{del}.
\begin{figure}
\includegraphics[width=1.\linewidth]{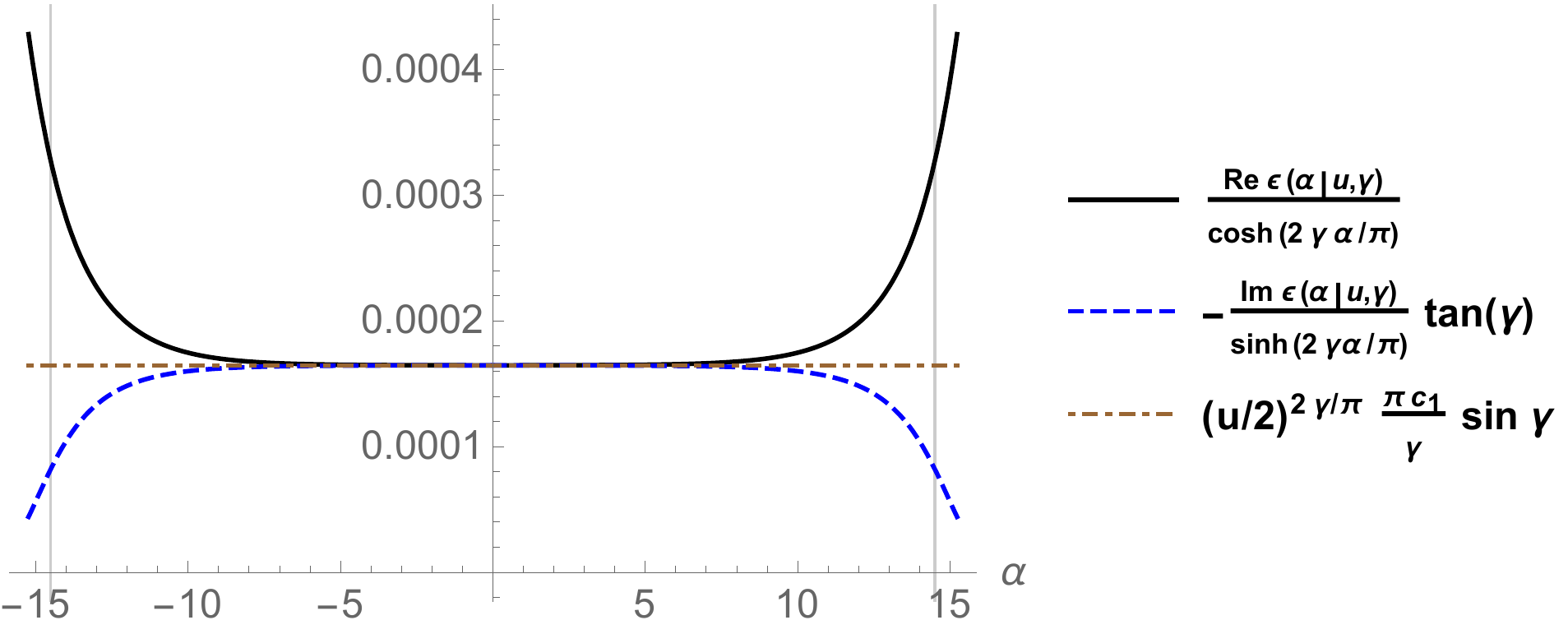}
\caption{Numerical check of equations \eref{a}, \eref{ve4}, \eref{ReE},  \eref{rr}, and \eref{eq:c1} at $\gamma=0.3\pi$, $u=10^{-6}$. At a small value of $u$, the functions ${\mathrm{Re}}\,{\varepsilon}(\alpha|u,\gamma)/\cosh (2 \gamma\alpha/\pi)$ 
and $- (\tan \gamma)\, {\mathrm{Im}}\,{\varepsilon}(\alpha|u,\gamma)/\sinh (2 \gamma\alpha/\pi)$ 
have wide plateaus near the origin $\alpha=0$, where they  approach almost the same  value. 
At $u\to0$ this value is given by equations \eref{ReE}, \eref{eq:c1}.
Vertical lines are located at  
$\pm \ln(2/u)$.}
\label{ratio}
\end{figure}

\begin{figure}
\centering
\subfloat[ ]
{\label{Ycas}
\includegraphics[width=1.\linewidth]{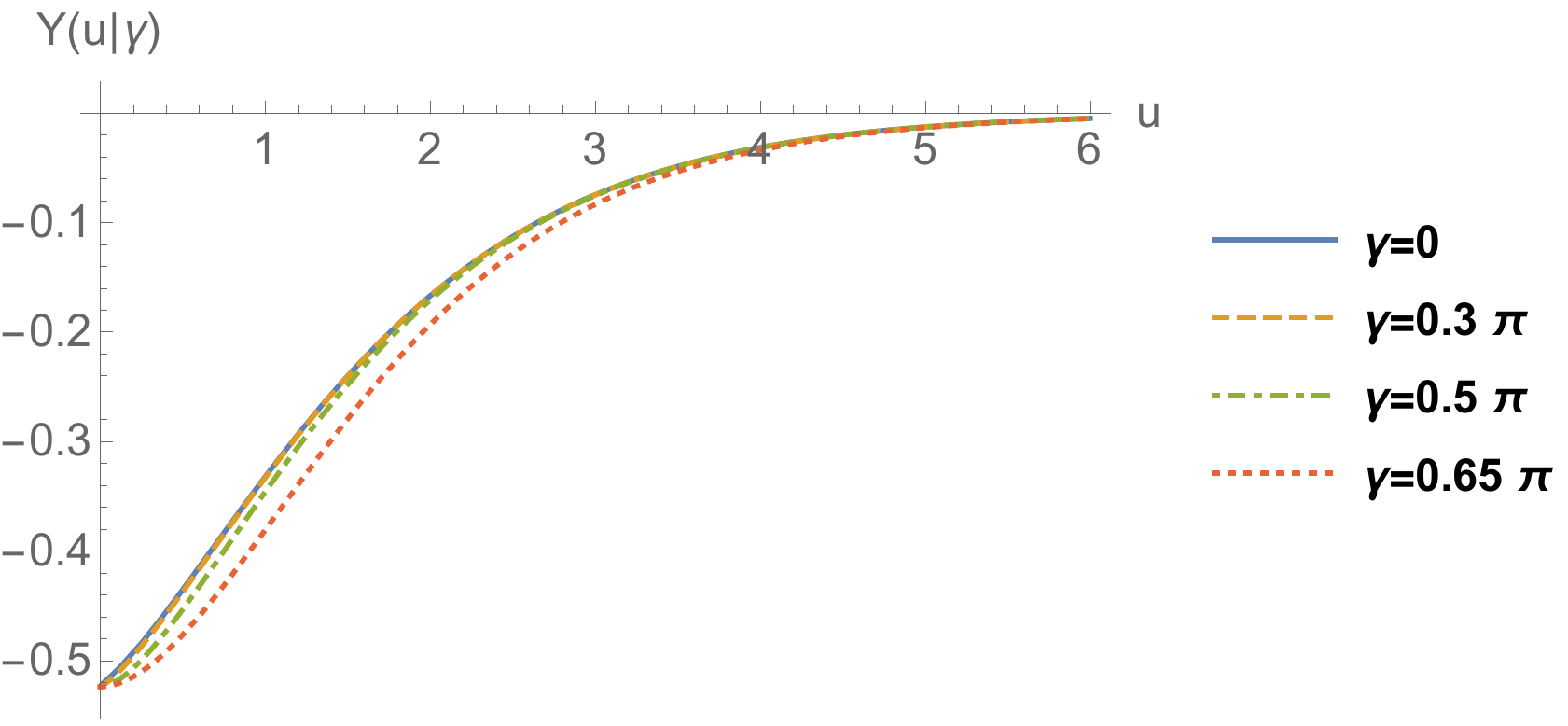}}

\subfloat[]{
	\label{ddY}
\includegraphics[width=1.\linewidth]{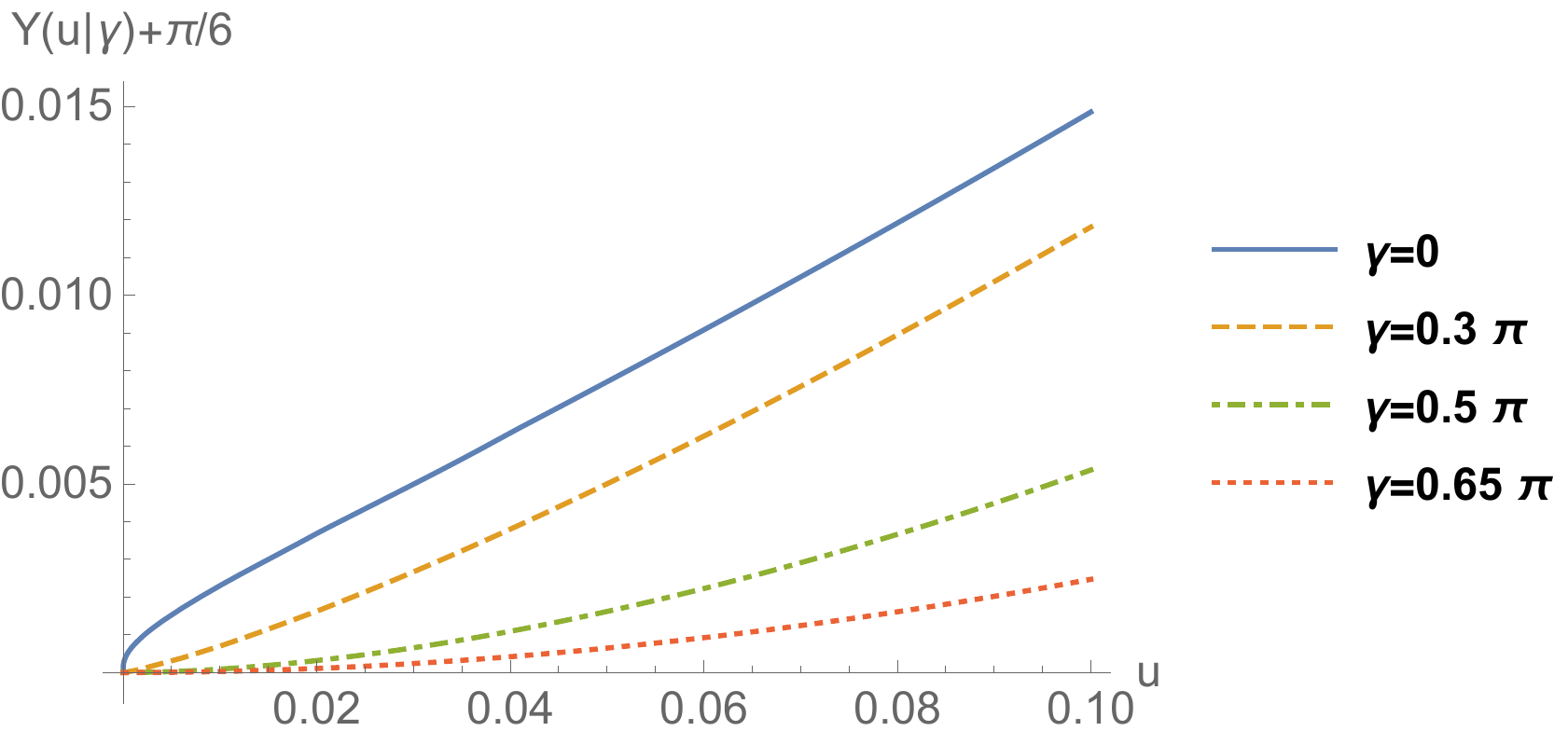}}

\caption{(a) Casimir scaling function $Y(u|\gamma)$, and (b) its deviation from the CFT value $-\frac{\pi}{6}$,
plotted against the scaling parameter $u$  at different values of $\gamma$.}
\label{4}
\end{figure}

\begin{figure}
\centering
\subfloat[ ]
{\label{Ycs}
\includegraphics[width=1.\linewidth]{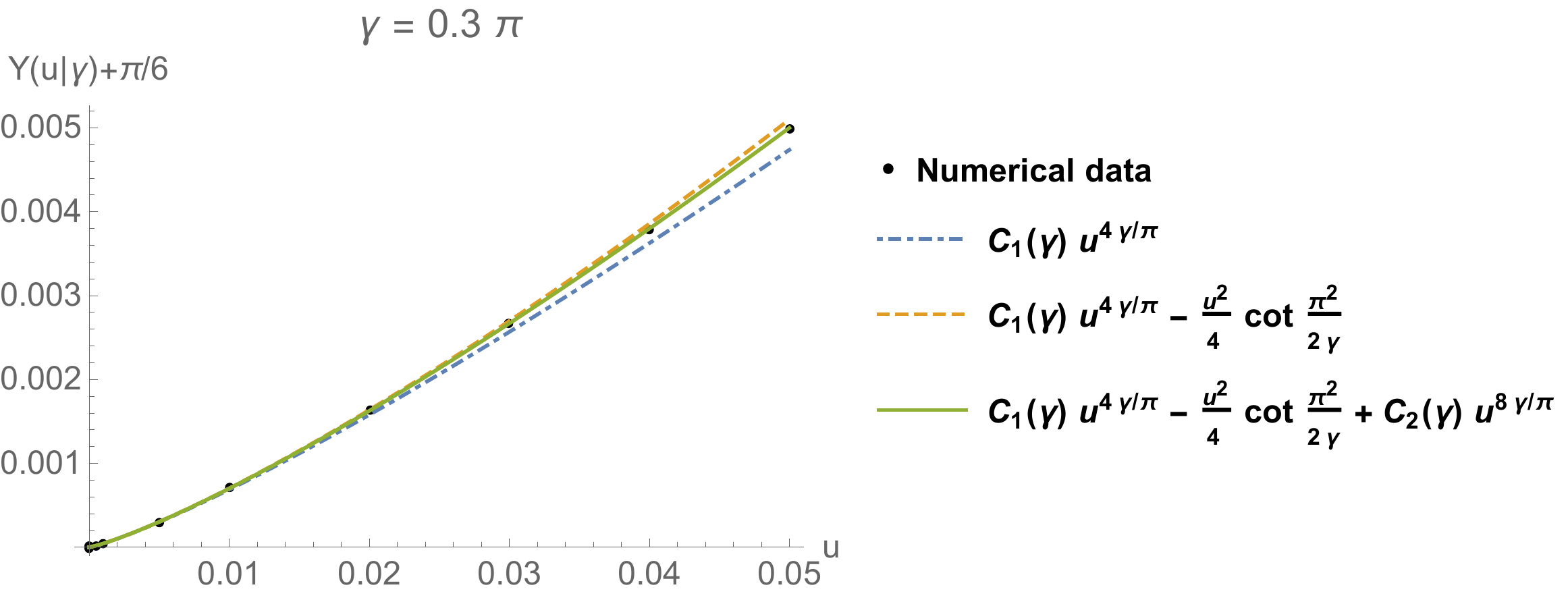}}

\vspace{.1cm}
\subfloat[]{
	\label{ddY1}
\includegraphics[width=1.\linewidth]{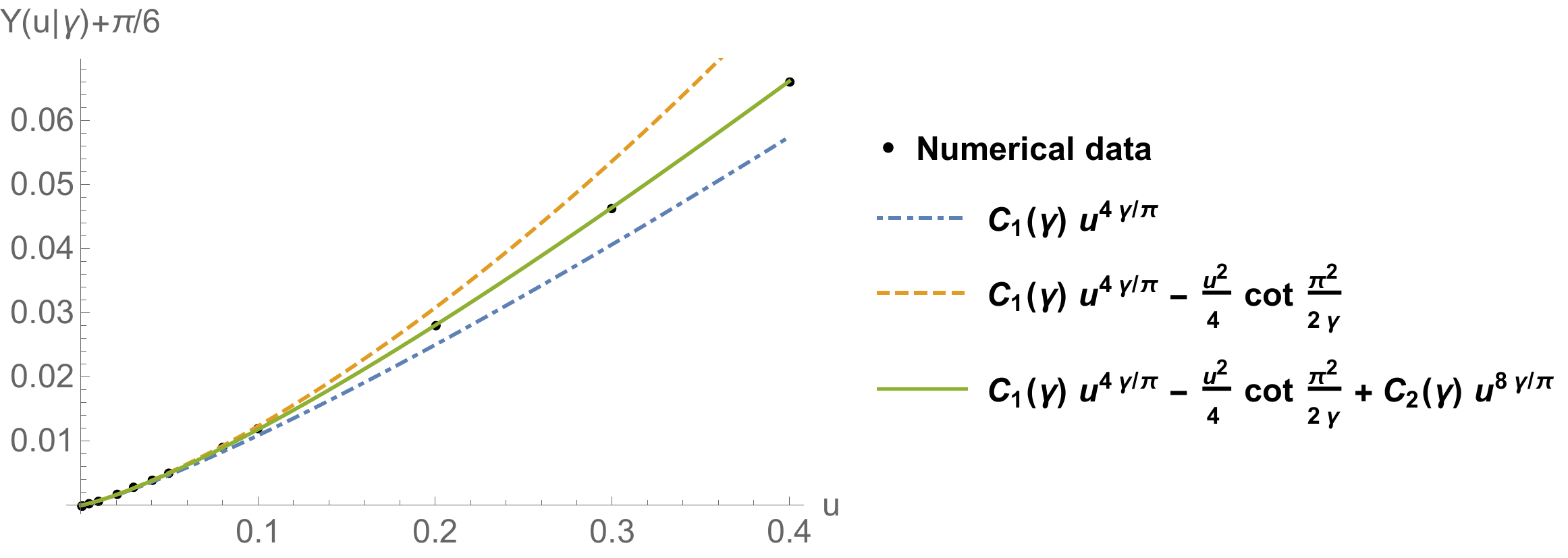}}
\caption{Casimir scaling function $Y(u|\gamma)+\frac{\pi}{6}$ in the repulsive regime at $\gamma=0.3\pi$ plotted against the scaling parameter $u$:   (a) at $0<u<0.05$, (b) at $0<u<0.4$. Numerical data are shown by dots, 
The coefficient $C_1(\gamma)$ is given by  \eref{C1c0}, the coefficient $C_2(0.3\pi)=-0.13$ was obtained by fitting 
the numerical data in the interval $0<u<0.05$.}
\label{sm}
\end{figure}

Figure \ref{dev} displays the  real and imaginary parts 
of the difference 
\begin{equation}\label{delte}
\Delta \varepsilon(\alpha|u,\gamma)=\varepsilon(\alpha|u,\gamma)-u \cosh \alpha
\end{equation}
plotted agains the rapidity $\alpha$ in the repulsive regime at $\gamma=0.3 \pi$ and $u = 10^{-6}$. 
It was shown in the previous Section, that the pseudoenergy  $\varepsilon_k(\alpha|\gamma)$ corresponding 
to the critical point $u=0$ decays at $\alpha\to-\infty$  as $\sim e^{2\gamma\alpha/\pi}$. 
Despite the 
claim of Destri and de Vega in  \cite{De94},  this exponential decay is slower than $e^\alpha$
in the repulsive regime $0<\gamma<\pi/2$. By this reason, and due to the asymptotical formulas
\eref{ve3}, \eref{ve4}, the first term in the right-hand side of \eref{delte} dominates over
the second one 
at $R(u)-|\alpha|\gg1$. Therefore, 
\[
|\Delta\varepsilon(\alpha|u,\gamma)|\gg u \cosh \alpha
\]
 in this region of 
the rapidity variable $\alpha$.  As one can see in Figure \ref{dev}, the above strong inequality really holds 
at $|\alpha|\lesssim10$ for the chosen values of parameters $\gamma,u$.

Figure \ref{ratio} provides a more detailed numerical check of the small-$u$ asymptotical 
formulae  \eref{ve4}, \eref{a}, \eref{RI}, and \eref{rr} for the pseudoenergy $\varepsilon(\alpha|u,\gamma)$ at $\gamma=0.3\pi$, and $u=10^{-6}$. The solid black and the  dashed blue lines display the numerically obtained $\alpha$-dependences
of the functions ${\mathrm{Re}}\,{\varepsilon}(\alpha|u,\gamma)/\cosh (2 \gamma\alpha/\pi)$ 
and $- (\tan \gamma)\, {\mathrm{Im}}\,{\varepsilon}(\alpha|u,\gamma)/\sinh (2 \gamma\alpha/\pi)$, respectively. In a
agreement with the theoretical predictions \eref{RI}, \eref{rr},    these two functions display wide plateaus near the origin $\alpha=0$
approaching there to almost the same value. The dot-dashed brown horizontal line represents the theoretical prediction  \eref{ReE}
 for this  value at $u\to 0$, with the
constant $c_1(\gamma)$ given by \eref{eq:c1}. 

Figure \ref{4} summarises the numerical results for the $u$-dependences of the Casimir scaling function 
$Y(u|\gamma)$ at four different values of parameter $\gamma$. The blue solid lines corresponding to 
the degenerate repulsive regime $\gamma=0$ were plotted using the previously obtained results 
\cite{Rut20}. The orange dashed lines correspond to the repulsive case at $\gamma=0.3\pi$. The dashed 
orange line cannot be distinguished from the solid blue one in Figure~\ref{Ycas} showing  the scaling function  variation 
in the wide interval of the scaling parameter $0<u<6$. However, the solid blue and the dashed orange lines are 
well separated in Figure \ref{ddY} that displays the scaling function variation in the small-$u$ region
$0<u<0.1$. The green dot-dashed lines in Figure \ref{4} display the scaling function \eref{frf}
at the free-fermionic point. The red dot-lines  in Figure  \ref{4} display   the Casimir scaling function  in the attractive regime
$Y(u|0.65\pi)$. 
\begin{figure}[t]
\includegraphics[width=1.\linewidth]{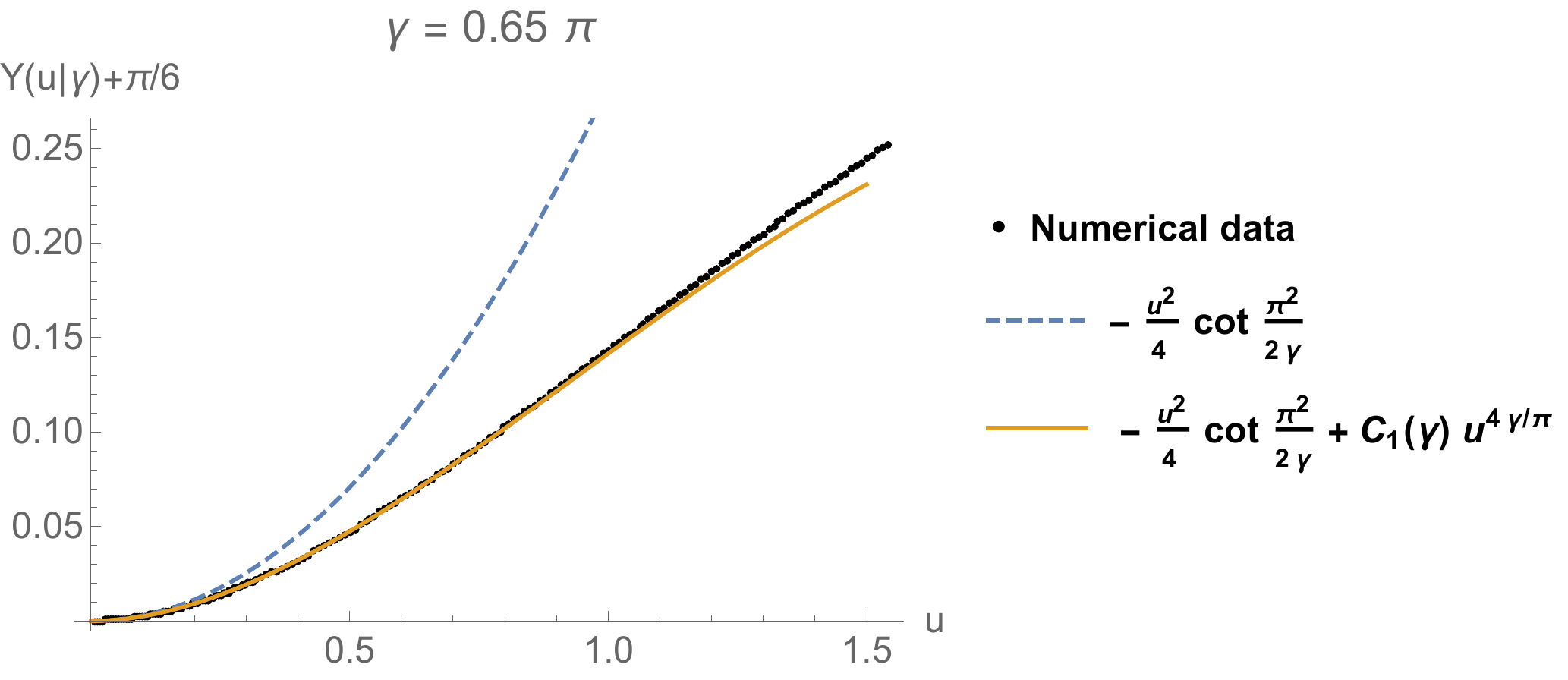}
\caption{Casimir scaling function $Y(u|\gamma)+\frac{\pi}{6}$ in the attractive regime at $\gamma=0.65 \pi$ plotted against the scaling parameter $u$ at $0<u<1.5$. Numerical data are shown by dots. The dashed blue line displays the leading term
in the small-$u$ asymptotics \eref{Y} of the scaling function. The  solid  orange line shows the plot of the two initial terms in the asymptotical
formula \eref{Y} with the coefficient $C_1(\gamma)$  given by  \eref{C1c0}.}
\label{de1}
\end{figure}

Figure \ref{sm} shows 
the $u$-dependence of the deviation of the Casimir scaling function $Y(u|\gamma)$ from its CFT value $-\pi/6$ 
in the repulsive regime at $\gamma=0.3 \pi$ in the small-$u$ regions, $0<u<0.05$ in Figure \ref{Ycs}, 
and $0<u<0.4$ in Figure \ref{ddY1}. The dots display the numerical data. The dot-dashed blue, dashed orange and 
solid green lines
display one, two and three leading terms, respectively,  in the asymptotical formula \eref{Y}. The
coefficient $C_1(\gamma)$ is given explicitly by \eref{C1c0}. The analytical expression for the coefficient
$C_2(\gamma)$ is not known.  We obtained the numerical value $C_2(\gamma=0.3 \pi)\approx -0.13$ 
of this coefficient 
by fitting the numerical results for the sum
\[
Y(u|\gamma)+\frac{\pi}{6}-C_1(\gamma)\, u^{4\gamma/\pi}+\frac{u^2}{4} \cot\left(\frac{\pi^2}{2 \gamma}\right)
\]
 in the interval $0<u<0.05$.

Figure \ref{de1} displays the scaling function $Y(u|\gamma)+\pi/6$ in the attractive regime at 
$\gamma=0.65 \pi$ in the  region $0<u<1.5$. The black dots show the numerical data, the blue dashed line
plots just one leading term in the small-$u$ asymptotic expansion \eref{Y}, and the solid orange line displays
the sum of two leading terms in this expansion. Note, that at $\gamma=0.65 \pi$  only two initial terms in the 
small-$u$ asymptotic expansion  \eref{Y}
describe remarkably   accurate the numerical data for the Casimir scaling function in a very wide interval of the 
scaling parameter $u$. 
\ack A am thankful to Frank G\"ohmann  and   {\color{black} Andreas Kl\"umper} for helpful discussions.
\appendix
\section{\label{PTBA} Derivation of \eref{Yas} from the TBA equation}

In this Appendix, we perform the  asymptotic analysis of the nonlinear TBA integral equations \eref{tBA1}  at a small 
$u\to0$, and obtain formula \eref{Yas} for the Casimir scaling function \eref{Ysc} in the repulsive regime $\gamma\in(0,\pi/2)$. The calculation is based on the technique developed in  \cite{Rut20}.

Using the reflection symmetry \eref{refl} of the  function ${\varepsilon}(\alpha|u,\gamma)$
and partial integration, we rewrite the 
integral formula \eref{Ysc}  for the scaling function in the form
\begin{equation}\label{YscA}
\fl Y(u|\gamma) =
- \frac{\,u}{\pi} \,{\rm Re}\int_{-\infty}^\infty {d\alpha}\, e^{\alpha}\,  L(\alpha|u,\gamma)
=  \frac{\,u}{\pi} \,{\rm Re}\int_{-\infty}^\infty {d\alpha}\, e^{\alpha}\, \partial_\alpha\,  L(\alpha|u,\gamma).
\end{equation}
Then we divide the integral $ \int_{-\infty}^\infty {d\alpha}$ on the right-hand side into two parts, 
$ \int_{-\infty}^\infty {d\alpha}=\int_{-\infty}^{0} {d\alpha}+\int_{0}^\infty {d\alpha}$.  The first 
term is small $\sim u^{\color{black}1+2\gamma/\pi}$.
{\color{black}Omitting} this term, one obtains at $u\to 0$,
\begin{equation}\label{Yu1}
Y(u) =\frac{2}{\pi}\, {\rm Re} \int_{0}^\infty {d\alpha}\,\frac{u\,e^\alpha}{2}\, 
\partial_\alpha\,L(\alpha|u,\gamma)
+O(u^{\color{black}1+2\gamma/\pi}).
\end{equation}

Let us now rewrite the TBA equation \eref{tBA3} as,
\begin{eqnarray}
&\frac{u\, e^\alpha}{2}=\varepsilon(\alpha|u,\gamma){\color{black}-} \frac{u \,e^{-\alpha}}{2}
+2  \int_{-\infty}^{\infty}
d\alpha' \, G(\alpha-\alpha'|\gamma)\,{\rm Re}\, L(\alpha'|u,\gamma) \\
 &+\int_{-\infty}^{\infty}d\alpha' \, g(\alpha-\alpha'- i 0|\gamma) \,L(\alpha'|u,\gamma),\nonumber
 \end{eqnarray}
substitute its  right-hand side instead of the factor $u \,e^\alpha/{2}$ into the integrand in \eref{Yu1},
and perform the term-wise  integration of the resulting integral. 

In the first term, the integration
can be performed explicitly:
\begin{eqnarray}\nonumber
&J_1(u)\equiv-\frac{2}{\pi}\, {\rm Re} \int_{0}^\infty {d\alpha}\,\varepsilon(\alpha|u,\gamma)\, 
\frac{\partial_\alpha\, \varepsilon(\alpha|u,\gamma) }{1+\exp[\varepsilon(\alpha|u,\gamma)]}=\\
&-\frac{2}{\pi} \int_{\varepsilon(0|u,\gamma)}^\infty  dx \,\frac{x}{1+e^x}=
-\frac{\pi}{6}+\frac{[\varepsilon(0|u,\gamma)]^2}{2\pi}+O([\varepsilon(0|u,\gamma)]^3).\label{J1}
\end{eqnarray}
We use  representation  \eref{ve1} for the pseudoenergy  $\varepsilon(0|u,\gamma)$, and drop  
in it the correction term  
$\eta{(0|u,\gamma)}$, which does not contribute  in \eref{J1} to the leading order in $u$. 
Recalling, that $R(u)=\ln(2/u)$ and using expansion \eref{vek}, one obtains then:
\begin{eqnarray}\nonumber
&\varepsilon(0|u,\gamma)\cong 2\, {\rm Re}\, \, \varepsilon_k\!\left(-R(u)|u,\gamma\right)
=
\pi c_1(\gamma)\,\frac{\sin \gamma}{\gamma}\,
\left(
\frac{u}{2}\right)^{2\gamma/\pi}+O(u^{4 \gamma/\pi}),\\
 &J_1(u)=-\frac{\pi}{6}+\frac{\pi}{2 }\label{J1}
\left(\frac{ c_1(\gamma)\sin \gamma}{\gamma}\right)^2\,
\left(
\frac{u}{2}
\right)^{4\gamma/\pi}+O(u^{6\gamma/\pi}).
\end{eqnarray}
The second integral 
 \[
 J_2(u)\equiv{\color{black}-} \frac{2}{\pi}\, {\rm Re} \int_{0}^\infty {d\alpha}\,\frac{u \,e^{-\alpha}}{2}\,\partial_\alpha\,L(\alpha|u,\gamma)
 \]
will be dropped, since it  vanishes  ${\color{black}\sim u^ {1+2\gamma/\pi}}$  at $u\to0$. 

The sum of the two remaining double-integrals
\begin{eqnarray}\nonumber
 & J_3(u)\equiv \frac{2}{\pi}\, {\rm Re}\Bigg\{\int_{0}^\infty {d\alpha}\,\partial_\alpha\,L(\alpha|u,\gamma)
  \int_{-\infty}^{\infty} d\alpha' \Big[2 G(\alpha-\alpha'|\gamma)\,{\rm Re}\, L(\alpha'|u,\gamma) \\
 & +g(\alpha-\alpha'- i 0|\gamma) \,L(\alpha'|u,\gamma) \Big]\Bigg\}\label{QQ}
\end{eqnarray} 
can be represented as the sum of three terms:
\begin{equation}\label{J3}
J_3(u)=A+B+C,
\end{equation}
where
\begin{eqnarray}\label{ABC}
&A=\frac{2}{\pi }[L(0|u,\gamma)]^2\,{\rm Re} \int_{2 R(u)}^\infty d\alpha\, U(\alpha),\\
&B=-\frac{2}{\pi }L(0|u,\gamma)\,{\rm Re} \int_0^\infty d\alpha\, U(\alpha+R(u)) \Psi(\alpha|u),\\
&C=\frac{4}{\pi }\iint_0^\infty d\alpha d\alpha'\, {G}(\alpha+\alpha'|\gamma) {\rm Re}\, [\partial_\alpha L(\alpha|u)]{\rm Re}\, [\Psi(\alpha'|u)]+\nonumber\\\label{C}
&\frac{2}{\pi }{\rm Re}\iint_0^\infty d\alpha d\alpha' g(\alpha+\alpha'-i0|\gamma)\partial_\alpha L(\alpha|u,\gamma)\Psi(\alpha'|u),
\end{eqnarray}
and 
\begin{eqnarray}
&U(\alpha)=2 {G}(\alpha|\gamma)+g(\alpha|\gamma),\\
&\Psi(\alpha|u)=L(\alpha|u,\gamma)-\Theta\big(R(u)-\alpha\big)\,L(0|u,\gamma).
\end{eqnarray}
Here $\Theta(x)$ denotes the unit-step function. Representation \eref{J3} - \eref{C} for the integral $J_3(u)$
defined by \eref{QQ}
is exact. In the degenerate case $\gamma=0$, it was obtained in  \cite{Rut20}, see equations (A31), (A32) there. 

We  substitute the function $\varepsilon(\alpha|u,\gamma)$ in the form  
 \eref{ve1} into the integrals in \eref{ABC}, expand the result in appropriate 
 fractional powers of $u$ using \eref{vek}, and keep only the leading term $\sim u^{4\gamma/\pi}$.
 It turns out, that: 
 \begin{enumerate}
 \item The correction term $\eta{(\alpha|u,\gamma)}$ in  the right-hand side of \eref{ve1} does not contribute to the above integrals  to this order.
\item  The following estimates $A\sim u^{4\gamma/(\pi-\gamma)}$, $B\sim u^{2\gamma/\pi+2\gamma/(\pi-\gamma)}$, 
$C\sim u^{4\gamma/\pi}$ hold for these integrals. Therefore,
in the considered regime $0<\gamma<\pi/2$, the integrals $A$ and $B$ can be dropped in \eref{J3}
to the leading order in $u$.
\item The main contribution into the double-integrals in \eref{C} comes from finite 
$\alpha, \alpha'\sim1$. At finite $\alpha,\alpha'\ll R(u)$, the following substitutions
can be safely made in the right-hand side of \eref{C}:
\[
\fl \partial_\alpha L(\alpha|u,\gamma)= -\frac{\partial_\alpha\varepsilon(\alpha|u,\gamma)}{2}+O(u^{4\gamma/\pi}),\quad
\Psi(\alpha|u)= -\frac{\varepsilon(\alpha|u,\gamma)}{2}+O(u^{4\gamma/\pi}),
\]
\[
\fl \varepsilon(\alpha|u,\gamma)=\frac{\pi i }{2 \gamma}\,c_1(\gamma)\left(
\frac{u}{2}
\right)^{2\gamma/\pi}\!\left[
\exp\!\left(
\frac{2\gamma\alpha}{\pi}+i\gamma
\right)-\exp\!\left(-
\frac{2\gamma\alpha}{\pi}-i\gamma
\right)
\right]+O(u^{4\gamma/\pi}).
\]
The resulting double-integrals in \eref{C} can be easily calculated using equalities \eref{equ}. This 
yields:
\begin{equation}\label{Cf}
C=\frac{\pi c_1(\gamma)^2}{2\gamma^2}
\left(
\frac{\pi}{2\sin 2 \gamma}-  \sin^2 \gamma
\right).
\end{equation}
\end{enumerate}
Combining \eref{Cf} with \eref{J1} and \eref{J3}, we arrive at the final result  \eref{Yas}.
\section{ \label{CFT} Perturbative CFT calculation of $Y(u)$}

The free energy $f(\mu,L)$ per unit area of the sine-Gordon model in the infinite strip of width $L$ is defined 
as the limit  
\begin{equation}
f(\mu,\gamma,L)=-\lim_{L'\to\infty}\frac{\ln Z(\mu,\gamma,L,L')]}{L\,L'},
\end{equation}
where the partition function $Z(\mu,L,L')$ on the torus ${\mathcal T}$ is determined by the continual integral \eref{pf}. Expansion of 
this integral to the second order  in $\mu$ yields 
\begin{eqnarray}
&-f(\mu,\gamma,L) =-f(0,\gamma,L)+2\mu \langle \cos[\beta\,\varphi(0)]\rangle+\\
&2\mu^2 \int_{-\infty}^{\infty}dx^0 \int_0^{L}dx^1
\langle \cos[\beta\,\varphi(x)]\cos[\beta\,\varphi(0)]\rangle\,\Theta(|x|-a)+O(\mu^4), \nonumber
\end{eqnarray}
where the correlation functions are connected and calculated at $\mu=0$. The cut-off at the lattice constant
$a$ serves to regularise the integral at  short distances. 
The linear in $\mu$ term in the right-hand side vanishes, since 
$
\langle \cos[\beta\,\varphi(0)]\rangle=0
$. The second-order  term $\delta_2 f(\mu,L)$ can be written as 
\begin{equation}\label{dfb}
 \fl \delta_2 f(\mu,\gamma, L)=-\mu^2 \!\!\int_{-\infty}^{\infty}\!\!\!dx^0\!\! \int_0^{L}\!\!\!dx^1
\langle \exp[i \beta\varphi(x)]
\exp[-i \beta\,\varphi(0)]\rangle\,\Theta(|x|-a).
\end{equation}
For the strip geometry, the correlation function  of the exponential operators in the integrand
can be determined \cite{Cardy_1984} by use of the conformal invariance at $\mu=0$,
\begin{equation}
\langle e^{i\beta \varphi(x)}e^{-i\beta \varphi(0)}\rangle=\left(\frac{2\pi}{L}\right)^{2{\mathrm x}}
\left\{
2[\cosh(2\pi x^0/L) -\cos(2\pi x^1/L)
]
\right\}^{-{\mathrm x}},
\end{equation}
with the scaling dimension ${\mathrm x}$ given by \eref{scd}. After substitution of this expression
into \eref{dfb} and rescaling  the integration variables, one finds,
\begin{equation}\label{df}
\delta_2 f(\mu,\gamma, L)=-\mu^2\left(\frac{2\pi}{L}\right)^{2{\mathrm x}-2}\,I_2({\mathrm x},\varepsilon),
\end{equation}
where
\begin{equation}\label{intI2}
I_2({\mathrm x},\varepsilon)= \int_{-\infty}^{\infty}dt \int_0^{2 \pi}d\theta \,
\frac{1}
{\left[2(\cosh
 t-\cos\theta
)
\right]^{{\mathrm x}}}\,\Theta\left(\sqrt{t^2+\theta^2}-\varepsilon\right),
\end{equation}
and $\varepsilon=2 \pi a/L$.

At $0<{\mathrm x}<1$, the parameter $\beta^2$ lies in the interval $(0,1/2)$, 
and  the attractive regime of the sine-Gordon model is realised. The integral $I_2({\mathrm x},0)$ converges in this case. It's explicit expression
was obtained by Hentschke {\it {et al.} }\cite{Hentschke_1986},
\begin{equation}\label{I2}
I_2({\mathrm x},0)=\pi\,\frac{\Gamma^2(\frac{1}{2}{\mathrm x})\,\Gamma(1-{\mathrm x})}
{\Gamma^2(1-\frac{1}{2}{\mathrm x})\,\Gamma({\mathrm x})}.
\end{equation}
At a small $\varepsilon>0$, the leading behaviour of the integral \eref{intI2} reads \cite{Cardy86},
\begin{equation}\label{I2a}
I_2({\mathrm x},\varepsilon)=I_2({\mathrm x},0)-\frac{\pi \varepsilon^{2-2{\mathrm x}}}{1-{\mathrm x}}+
O(\varepsilon^{6-2{\mathrm x}}).
\end{equation}
Equations \eref{df}, \eref{I2}, \eref{I2a} can be used 
in both $0<{\mathrm x}<1$ and $1<{\mathrm x}<2$ cases,  which correspond to the attractive 
and repulsive regimes of the sine-Gordon model, respectively. Substitution of \eref{I2a} into \eref{df} yields,
\begin{equation}\label{dfa}
\delta_2 f(\mu,\gamma, L)=-\mu^2\left(\frac{2\pi}{L}\right)^{2{\mathrm x}-2}\,I_2({\mathrm x},0)+
\frac{\pi \mu^2}{1-{\mathrm x}}\, a^{2-2{\mathrm x}}.
\end{equation}
In the repulsive regime $1<{\mathrm x}<2$, the $L$-independent term in the right-hand side  diverges 
at $a\to0$ and contributes to the non-universal part of the bulk free energy. 

Let us now use equation \eref{mum} to express the coupling constant $\mu$ in equation \eref{dfa} in terms of the soliton mass 
$m$,
\begin{equation}\label{df1}
\fl \delta_2 f[\mu(m),\gamma, L]=-\kappa(\xi)^2  m^{4/(\xi+1)}\left[\left(\frac{2\pi}{L}\right)^{2{\mathrm x}-2}\,I_2({\mathrm x},0)-\frac{\pi }{1-{\mathrm x}}\, a^{2-2{\mathrm x}}
\right].
\end{equation}
After replacement of parameters $\xi$ and ${\mathrm x}$ in this equation by their expressions in terms of the parameter $\gamma$, 
\[
\xi=\frac{\pi-\gamma}{\gamma}, \quad {\mathrm x}=\frac{2(\pi-\gamma)}{\pi},
\]
one obtains
\begin{equation}\label{df2}
\delta_2 f[\mu(m),\gamma, L]= \frac{C_1(\gamma)}{L^2} u^{4\gamma/\pi}-\frac{\pi^2}{(\pi-2 \gamma)a^2}
(m a)^{4\gamma/\pi},
\end{equation}
where $u=m L$ is the scaling parameter, and $C_1(\gamma)$ is given by \eref{C1c0}.

Let us recall now that due to equation \eref{ZZ} the ground-state energy $E(\mu,\gamma,L)$
of the sine-Gordon model Hamiltonian in the circle of length $L$ is proportional to the free energy
$f(\mu,\gamma, L)$ per unit area,
\begin{equation}
E(\mu,\gamma,L)=L\,f(\mu,\gamma, L).
\end{equation}
Combining this with equality \eref{EY} allows one to relate the Casimir scaling function
$Y(u|\gamma)$ with $f(\mu,\gamma, L)$ and the bulk energy density ${\mathcal E}_b(\mu,\gamma)$.
\begin{equation}\label{Yu}
Y(u|\gamma)=L^2\,[f(\mu,\gamma, L)-{\mathcal E}_b(\mu,\gamma)],
\end{equation}
The energy density of the slab $f(\mu,\gamma, L)$, as well as the bulk energy density  in the sine-Gordon model
\begin{equation}
{\mathcal E}_b(\mu,\gamma)=\lim_{L\to\infty}f(\mu,\gamma, L)
\end{equation}
contain the ultraviolet-diverging parts, and the (independent of the
 lattice spacing $a$) scaling parts. The explicit expression of the scaling part 
 ${\mathcal E}_b^{(sc)}[\mu(m),\gamma]$ of the bulk energy density is well known \cite{Bax72,De94}, \begin{equation}
{\mathcal E}_b^{(sc)}[\mu(m),\gamma]
=
\cases
{\frac{m^2}{4}\, \cot\left(\frac{\pi^2}{2 \gamma}\right),  \;{\mathrm{if}}\; \gamma\ne \frac{\pi}{2n}, \\
\frac{m^2}{2\pi n}\, \ln m, \quad \quad\;
\:{\mathrm{if}}\; \gamma= \frac{\pi}{2n},}
\end{equation}
where  $m$ is the soliton mass \eref{mum}, and $n=1,2,\ldots$. 
Since the universal Casimir scaling function $Y(u|\gamma)$ in the left-hand side of \eref{Yu} 
is ultraviolet-convergent, the same is true for the right-hand side of equation \eref{Yu}.
This implies, in particular, that the second ($L$-independent, ultraviolet-diverging) term in the right-hand side
of \eref{df2} must cancel with the same contribution from the bulk energy density 
${\mathcal E}_b(\mu,\gamma)$ in  brackets in the right-hand-side of \eref{Yu}.
So, we can replace the functions $f(\mu,\gamma, L)$ and ${\mathcal E}_b(\mu,\gamma)$ in 
the right-hand side of  \eref{Yu} by their scaling counterparts
\begin{eqnarray}\label{fesc}
&&f^{(sc)}(u,\gamma, L)=\frac{1}{L^2}
\left[
-\frac{\pi}{6}+C_1(\gamma)\,u^{4\gamma/\pi}+O(u^{8\gamma/\pi})
\right],\;{\mathrm{if}}\: \gamma\ne \frac{\pi}{2},\\
&&{\mathcal E}_b^{(sc)}(u,\gamma,L)
=
\cases
{\frac{u^2}{4 L^2}\, \cot\left(\frac{\pi^2}{2 \gamma}\right), \quad {\mathrm{if}}\; \gamma\ne \frac{\pi}{2n},
\\
 \frac{u^2}{2\pi n\,L^2} \, \ln u, \quad \quad \;\; {\mathrm{if}}\; \gamma= \frac{\pi}{2n}
,}
\end{eqnarray}
with $n=1,2,\ldots$. This leads to the  results \eref{Y}, \eref{Y3}, that hold at $\gamma\ne \pi/2$. 

The Casimir scaling function degenerates in the free-fermionic case $\gamma=\pi/2$ to the form,
\begin{equation}\label{frf}
Y(u|\pi/2)=-\frac{u}{\pi} \int_{-\infty}^\infty {d\alpha}\,\cosh \alpha\, 
\ln\left[1+e^{-u\cosh \alpha}\right].
\end{equation}
Its behaviour at small $u\to0$ is described by equation \eref{YFF}.
\section{Calculation of the coefficient $c_1(\gamma)$ in the attractive regime $\pi/2<\gamma<\pi$\label{BLZ}}
The DDV integral equation (3.14) studied in \cite{BLZ97} relates to the massless case $\mu\to 0$ of the sine-Gordon model \eref{HSG} 
perturbed by the 
uniform gauge "magnetic field"  $p$, which is   coupled to the soliton charge. At $p=0$, this equation reduces to the form:
\begin{equation}\label{DDVml}
i \ln a(\theta)=2 M \cos \frac{\pi \xi}{2}\,\, e^\theta-2\int_{-\infty}^{\infty}
d\theta' \, G(\theta-\theta') \,{\rm Im}\,\ln[1+a(\theta'-i 0)],
\end{equation}
where the integral kernel $G(\theta)$ is given by \eref{Sc}, the constant $M$ reads
\begin{equation}
M=\frac{\Gamma\left(\frac{\xi}{2} \right)\Gamma\left(\frac{1}{2} -\frac{\xi}{2} \right)}{\sqrt{\pi}}\left(
\Gamma(1-\beta^2)
\right)^{1+\xi},
\end{equation}
and parameters $\xi$ and $\beta$ are related with $\gamma$ according to \eref{gam}.
Upon the shift of the rapidity variable 
\begin{equation}\label{shift}
\alpha =\theta+\ln\left(2 M \cos \frac{\pi \xi}{2}\right),
\end{equation}
 and  
the substitution 
\begin{equation}
a(\theta) =\frac{1}{{\mathfrak{f}}_k(\alpha)}, 
\end{equation}
equation \eref{DDVml} transforms to the form 
\eref{mles}. 

It was shown in \cite{BLZ97}, that the function $a(\theta)$ solving equation \eref{DDVml} admits in the
attractive case $0<\xi<1$ the following representation:
\begin{equation}
a(\theta)=\frac{A(\lambda q)}{A(\lambda q^{-1})},
\end{equation}
where $\lambda=\exp(\frac{\theta}{1+\xi})$, $q=\exp(\frac{i \pi \xi}{1+\xi})$, and $A(\lambda)$ is the 
vacuum eigenvalue of the operator $\mathbf{Q}_+(\lambda)$, which is the CFT analogue of the $Q$-matrix
introduced by  Baxter \cite{BAXTER19731,BAXTER197325,BAXTER197348,Bax}. The function 
$ A(\lambda)$ has remarkable analytical properties. It was shown in  \cite{BLZ97}, in particular, that $A(\lambda)$
is an entire function in the attractive regime $0<\xi<1$, that can be represented in this case by a convergent product
\begin{equation}
A(\lambda)=\prod_{k=0}^{\infty}\left(
1-\frac{\lambda^2}{\lambda_k^2}
\right),
\end{equation}
and all   zeroes $\lambda_k^2$ of $A(\lambda)$ are positive.  Accordingly, the function 
$\phi(\theta) =i \ln a(\theta)$ admits  at $\theta\to-\infty$ the converging Taylor expansion in $\lambda$:
\begin{equation}\label{Taphi}
\phi(\theta)=\sum_{n=1}^\infty b_n \lambda^{2n}
\end{equation}
with real coefficients $b_n$, and 
\begin{equation}
b_1=2 \sin\left(
\frac{2\pi \xi}{1+\xi} 
\right)\sum_{k=1}^\infty \,\lambda_k^{-2 }.
\end{equation}
The explicit formula for the infinite sum in the right-hand side can be gained from equations  (3.21),  (2.34) and (2.35) in
\cite{BLZ97}:
\begin{equation}
\sum_{k=1}^\infty \,\lambda_k^{-2 }=\Gamma(1-2\beta^2)[\Gamma(\beta^2)]^2.
\end{equation}
The resulting expression for the coefficient $b_1$ reads:
\begin{equation}\label{b1}
b_1=2 \sin\left(
\frac{2\pi \xi}{1+\xi} 
\right)\,\Gamma(1-2\beta^2)[\Gamma(\beta^2)]^2.
\end{equation}

On the other hand, the functions $\phi(\theta)$ and $\phi_k(\alpha)$ coincide after the shift \eref{shift} of the rapidity argument:
$
\phi(\theta) =\phi_k(\alpha)
$. Therefore, expansion \eref{vek1} represents the analytical continuation (in the parameter $\xi$) of the Taylor expansion \eref{Taphi} into the 
interval $\xi\in(1,+\infty)$, which  corresponds to the repulsive regime of the sine-Gordon model. This leads to the simple relation
between the coefficients of these two expansions:
\begin{equation}
c_n(\gamma)=\frac{2 n \gamma}{\pi }\, b_n\,\left[2 M\,  \cos\frac{\pi(\pi-\gamma) }{2\gamma}\right]^{-2 n \gamma/\pi}.
\end{equation}
Taking \eref{b1} into account, we 
arrive at the following  representation for the first coefficient
\begin{equation}
\fl c_1(\gamma)=-\frac{4  \gamma}{\pi }\, \sin\left(
2\gamma
\right)\,\Gamma\left(\frac{2\gamma}{\pi}-1\right)\,\left[\Gamma\left(1-\frac{\gamma}{\pi}\right)\right]^2\,\left[2 M\, \cos\frac{\pi(\pi-\gamma) }{2\gamma}\right]^{-2  \gamma/\pi},
\end{equation}
which is equivalent to \eref{eq:c1}.

\end{document}